\documentclass[twocolumn,tighten,iop,times,twocolappendix]{aastex631}
\usepackage{graphicx}
\graphicspath{{Plots/}}
\usepackage{amsmath}
\usepackage{iondefs, farhan-defs}
\usepackage[super]{nth}

\usepackage{bigints}
\usepackage[fulladjust]{marginnote}

\usepackage{appendix}

\hypersetup{colorlinks=true, citecolor=blue, filecolor=magenta,urlcolor=blue, linkcolor=blue}

\begin{document}


\title{\sc Filaments of The Slime Mold Cosmic Web And How They Affect Galaxy Evolution}


\author[0000-0002-0072-0281]{Farhanul Hasan}
\affiliation{Department of Astronomy, New Mexico State University, Las Cruces, NM 88003, USA}

\author[0000-0002-1979-2197]{Joseph N. Burchett}
\affiliation{Department of Astronomy, New Mexico State University, Las Cruces, NM 88003, USA}

\author{Douglas Hellinger}
\affiliation{Department of Physics, University of California, Santa Cruz, CA 95064, USA}

\author[0000-0003-0549-3302]{Oskar Elek}
\affiliation{Department of Computational Media, University of California, 1156 High Street, Santa Cruz, CA 95064, USA}

\author[0000-0002-6766-5942]{Daisuke Nagai}
\affiliation{Department of Physics, Yale University, New Haven, CT 06520, USA}

\author[0000-0003-4996-214X]{S. M. Faber}
\affiliation{University of California Observatories and Department of Astronomy and Astrophysics, \\
University of California, Santa Cruz, 1156 High Street, Santa Cruz, CA 95064, USA}

\author[0000-0001-5091-5098]{Joel R. Primack}
\affiliation{Department of Physics, University of California, Santa Cruz, CA 95064, USA}

\author[0000-0003-3385-6799]{David C. Koo}
\affiliation{University of California Observatories and Department of Astronomy and Astrophysics, \\
University of California, Santa Cruz, 1156 High Street, Santa Cruz, CA 95064, USA}

\author[0000-0001-8057-5880]{Nir Mandelker}
\affiliation{Centre for Astrophysics and Planetary Science, Racah Institute of Physics, The Hebrew University, Jerusalem 91904, Israel}

\author{Joanna Woo}
\affiliation{Department of Physics, Simon Fraser University, 8888 University Drive, Burnaby, BC, V5A 1S6, Canada}



\shortauthors{Hasan et al.}

\begin{abstract}

We present a novel approach for identifying cosmic web filaments 
within the {\disperse} structure identification framework,
using cosmic density field estimates from the Monte Carlo Physarum Machine (MCPM), inspired by the {\it slime mold} organism.
We apply our method to the IllustrisTNG (TNG100) cosmological simulations and investigate the impact of filaments on galaxies. 
The MCPM density field is superior to the Delaunay Tessellation Field Estimator (DTFE) in tracing the true underlying matter distribution and allows filaments to be identified with higher fidelity, 
finding more low-prominence/diffuse filaments.
Using our new filament catalogs, we find that $\gtrsim$90\% of galaxies are located within $\sim$1.5 Mpc of a filamentary spine, with little change in the 
median star formation activity  
with distance to the nearest filament. 
Instead, we uncover a differential effect of the local filament line density, {\sigmafil} -- the total MCPM overdensity per unit length along a filament segment -- on galaxy formation:
most galaxies are quenched and gas-poor near high-line density filaments at $z\leq1$. At earlier times, the filamentary environment appears to have no effect on galactic gas supply and quenching.
At $z=0$, quenching in ${\logms}\gtrsim10.5$ galaxies is mainly driven by mass, while lower-mass galaxies are significantly affected by the filament line density. 
Satellites are far more susceptible to filaments than centrals. 
The local environments of massive halos are not sufficient to account for the effect of filament line density on gas removal and quenching. Our new approach holds great promise for observationally identifying filaments from galaxy surveys such as SDSS and DESI.

\end{abstract}


\section{Introduction}

Understanding how galaxies form and change over time is a major challenge in astrophysics. The mass of a galaxy plays a crucial role in determining its observable physical properties, as it determines the strength of internal processes such as supernovae (SNe) and active galactic nucleus (AGN) feedback \citep[e.g.,][]{Kauffmann04,Baldry06,Faber07,Peng10,Ilbert13,Schawinski14}. Moreover, the environment surrounding galaxies is also a significant factor in shaping their physical properties such as mass, highlighting the importance of the interaction between a galaxy and its surroundings in the process of evolution \citep[e.g.,][]{Dressler80,Balogh99,Elbaz07,Darvish16,Kraljic18,Sobral22}. 

Identifying the salient environmental indicator of a galaxy is a task that has been approached differently in various studies. Dark matter (DM) halos in which galaxies form play a critical role in shaping galactic properties \citep[e.g.,][]{Yang12,Behroozi13,Zu16,WT18,LSD23}. The environmental effects are particularly pronounced in satellite galaxies in massive clusters and groups \citep{Moore96,Poggianti06,Delucia07,Tonnesen07,Brown17,Zavala19}.
Therefore, it is expected that the evolution of galaxies depend on local environments typically characterized by the galaxy overdensity or the distance to the $N^{\mathrm{th}}$ nearest neighbor \citep[or variants thereof; e.g.,][]{Blanton05,Peng12,Woo13,Darvish14,Davidzon16,Moutard18}.

Galaxies can be analyzed in the context of the universe's large-scale structure, known as the "cosmic web." This structure consists of an interconnected network of filaments, which are bridges of intergalactic matter, and nodes, which are dense intersections of filaments where the cosmic density distribution is highest \citep[e.g.,][]{White87,Bond96,Springel05,disperse2,Cautun14,Martizzi19}. Galaxies are generally found within these structures. 
Owing to the anisotropic gravitational collapse of matter in the primordial density field, the cosmic web is intrinsically hierarchical and multi-scale; small structures merge to form larger structures over time \citep[e.g.,][]{Zeldovich70,Peebles80}. Within the cosmic web, filaments form on large scales of Megaparsecs -- which contain the majority of the baryons in the universe \citep[e.g.,][]{CO06,AC10,Snedden16,Martizzi19} and about half of the galaxies and DM halos \citep[e.g.,][]{Cautun14,Metuki15,GV19} -- and on much smaller scales, down to kiloparsecs -- which can penetrate galaxies as streams of cold gas \citep[e.g.,][]{DB06,Dekel09,Pichon11,Mandelker18,Mandelker20,Ramsoy21}. 
In this work, we focus on the large-scale filaments of physical size on the order of Mpc or larger to understand the impact of large-scale structure on the much smaller scales of galaxy formation, recognizing the diversity of physical processes at play on the different scales.

The cosmic web structure has been inferred from galaxy redshift surveys for several decades \citep[e.g.,][]{delapparent86,Geller89,Colless01,Strauss02,Lilly07,Driver11,Huchra12,Bryant15,Laigle16}. Although the nodes of the cosmic web can be easily identified by the location of the most massive halos, such as clusters and groups \citep[e.g.,][]{Kaiser84,ST04,Walker19,Kuchner20,Cohn22}, identifying the filamentary structure is much more complicated.

Identifying the filaments of the cosmic web from a sparse distribution of galaxies, commonly known as cosmic web ``reconstruction," is a challenging task. Various techniques have been proposed to achieve this goal, 
including graph theory and percolation-based methods, such as minimal spanning trees \citep[e.g.,][]{Alpaslan14a,Bonnaire20}, stochastic methods, such as the Bisous point process model \citep[e.g.,][]{Tempel14}, Hessian-based methods exploiting the geometry of local density, tidal, or velocity shear fields \citep[e.g.,][]{Cautun13}, phase-space methods based on the dynamics of the cosmic web \citep[e.g.,][]{Falck12}, manifold crawling and/or swarm intelligence-based methods \citep[e.g.,][]{Awad23}, and topological methods that exploit the connectivity and topology of space, such as {\sc Spineweb} \citep{AC10} and {\disperse} \citep{disperse1,disperse2}.
For a comprehensive comparison of these methods and the resulting cosmic web reconstructions, see \citet{Libeskind18}.  According to the authors, there is general agreement between the spatial and statistical distribution of the identified structures. Voids occupy most of the volume, and clusters live in nodes. However, there are significant disagreements in some cases, such as the mass/volume fractions in filaments. These disagreements can lead to different conclusions being drawn on the effect of large-scale structure on galaxy formation \citep{Rost20}.

Recent advances in cosmic web reconstruction techniques have enabled us to study the impact of filaments on various aspects of galaxy formation. These studies have explored the effect of filaments on quenching \citep[e.g.,][]{Kuutma:2017aa,Laigle18,Xu20,Winkel21,Bulichi23,Hasan23}, gas supply \citep[e.g.,][]{Stark16,kleiner17,CroneOdekon:2018aa,Song21,GE22,RG22,Zhu22}, and alignment between galaxy spin and filament direction \citep[e.g.,][]{Dubois14,Codis18,GV19,Pandya19,Kraljic19,bluebird20,Welker20,tudorache22}, in both observations and simulations. However, the methods used to identify filaments vary widely, and the conclusions drawn regarding the physical effect of filaments on galaxy formation phenomena are often divergent. For example, while some studies suggest a suppression of gas \citep[e.g.,][]{CroneOdekon:2018aa,Zhu22,Hasan23} and star formation \citep[e.g.,][]{Alpaslan16,Kraljic18,Malavasi22} near filaments, others report an increase in gas supply and/or star formation \citep[e.g.,][]{kleiner17,Vulcani19,Kotecha22}.

This work follows from \citet[][hereafter \citetalias{Hasan23}]{Hasan23}, who use the aforementioned {\disperse} framework to reconstruct the cosmic web in all snapshots of the IllustrisTNG cosmological simulations \citep{Pillepich18,Nelson18,TNGDR19} between redshifts $z=0$ and $z=5$. We examined the effect of proximity to the nearest cosmic web filaments and nodes on the quenching and gas supply of galaxies at different masses and epochs. This study revealed that, at earlier times ($z\gtrsim2$), the average star formation activity of galaxies is not affected by proximity to the cosmic web. However, at later times, galaxies are more likely to quench closer to nodes and filaments. This trend is particularly noticeable in low-mass galaxies ({\lowmsrange}) and satellites of all masses. This study also found that the primary reason behind these effects is the decline in galactic gas supply with cosmic web proximity.

In this paper, we present a new technique for extracting cosmic web structures from galaxy catalogs. We compare the impact of the resulting structure on galaxy evolution between our method and the one used in \citetalias{Hasan23}. Our approach uses a novel model called the Monte Carlo Physarum Machine \citep[MCPM; ][]{Burchett20,Elek21,Elek22} to estimate the cosmic density field. MCPM is inspired by the feeding habits of the biological organism {\it Physarum polycephalum} or {\it slime mold}, which is known to generate highly efficient interconnected networks when searching for food. This behavior has been used to model spatial problems in various disciplines, from neuroscience to civil engineering \citep[see, e.g., the review from][]{Adamatzky10}.

The MCPM algorithm takes a catalog of point sources as input, which in this work is a catalog of galaxies with stellar masses {\minms}. It then infers a continuous cosmic matter density field at every location in the input volume, {\it even in underdense regions far from galaxies.} This is what makes the MCPM density field distinct from other methods, such as the Delaunay Tessellation Field Estimator \citep[DTFE;][]{SV00,VS09}, which is integrated into {\disperse} and linearly interpolates density field estimates around the locations of galaxies. The MCPM model has been applied to analyze observational galaxy catalogs and fast radio burst data \citep{Burchett20,Simha20,Wilde23}. The MCPM density field has successfully recovered the density structure of the cosmic web and provided insights into the intergalactic 
neutral hydrogen gas \citep{Burchett20} and hot ionized gas \citep{Simha20} in the cosmic web.

In this work, we replace the standard DTFE density field with the MCPM density field in {\disperse} to reconstruct the cosmic web structure in six snapshots (ranging from $z=0$ to $z=4$) of IllustrisTNG (TNG100). 
Our approach has significantly improved the filament identification scheme in {\disperse}, possibly the most widely used cosmic web reconstruction tool.
We highlight the effect of cosmic web environment on the quenching and gas supply of galaxies using filaments from our new density field reconstruction. 
This finding yields invaluable predictions of the interplay between galaxies and large-scale structures, especially as our method is applied to observational galaxy catalogs.

The paper is organized as follows. In Section~\ref{sec:data}, we describe the simulation data and the methods used to reconstruct the cosmic web. We also compare the two different density field estimation methods of interest. In Section~\ref{sec:results}, we present our findings on how galactic star formation and gas fraction depend on cosmic web environments. We discuss our results in Section~\ref{sec:discuss}, and conclude the paper in Section~\ref{sec:conclusion}. We adopt {\it Planck 2015} cosmology \citep{Planck15}, as in \citetalias{Hasan23}, with $H_{0}=67.74$ {\kmsmpc}, $\Omega_{\mathrm{M},0} = 0.3089$, and $\Omega_{\Lambda,0} = 0.6911$. Unless otherwise stated, all distances and lengths are in comoving units.


\section{Data and Methods of Cosmic Web Reconstruction}
\label{sec:data}


\subsection{TNG Simulations}
\label{sec:tngdata}

We analyzed the outputs from the IllustrisTNG magneto-hydrodynamical cosmological simulations \citep{Pillepich18,Nelson18,TNGDR19}, which used the {\sc AREPO} moving-mesh hydrodynamic code \citep{Springel10} to simulate the evolution of gas, stars, dark matter, and black holes from the early universe ($z\!=\!127$) to the present day ($z\!=\!0$). Same as \citetalias{Hasan23}, we use the highest resolution run of the TNG100 simulation, TNG100-1, which has a box size of approximately 110.7 comoving Mpc per side, a minimum baryonic particle mass of $\sim\!1.4 \!\times\! 10^{6}~{\Msun}$, and cosmology based on {\it Planck 2015} \citep{Planck15}. 

We obtain galaxy data for all 100 snapshots of the TNG100-1 simulation (hereafter TNG) from the online data repository\footnote{\url{https://www.tng-project.org/data/}}, as presented first in \citet{TNGDR19}. In each snapshot, the ``Group" catalogs were constructed using the friends-of-friends (FoF) substructure identification algorithm, and the ``subhalo" catalogs were constructed using the {\sc Subfind} algorithm \citep{Springel01,Dolag09}. The ``subhalo" catalogs contain gravitationally bound objects within a FoF group, while the ``Group" catalogs identify halos, i.e., FoF groups. 
We extract data from both of these types of catalogs.
Hereafter, we refer to groups as halos and subhalos as galaxies.

Similarly to \citetalias{Hasan23}, we set a minimum stellar mass of $10^{8}~{\Msun}$ and a DM halo mass of $10^{9}~{\Msun}$ within a sphere of mean density 200 times the critical density of the Universe in each snapshot. This ensures that the galaxies are adequately resolved, having at least around 100 stellar particles, and correspondingly, their host halos have around 100 DM particles. 
We then extracted the following data from the galaxy and halo catalogs at each snapshot from $z=4$ to $z=0$: the galaxy's comoving position, its star formation rate (SFR), its stellar mass ({\ms}), its halo mass ({\mh}), its halo virial radius ({\Rh}; comoving radius where {\mh} is calculated), the mass of all gas gravitationally bound to a subhalo ({\mgas}), and synthetic stellar magnitudes in the SDSS $g,r,i,z$ bands \citep{Stoughton02}. For more information on the photometric bands, including the comprehensive dust attenuation modeling, see \citet{Nelson18}.
We also extracted DM particle data at each snapshot for several aspects of our study.

\vspace{-5pt}

\subsection{{\disperse} Cosmic Web Reconstruction}
\label{sec:disperse}

In this work, we use the Discrete Persistent Structures Extractor ({\disperse}) algorithm\footnote{\url{http://www2.iap.fr/users/sousbie/web/html/indexd41d.html}} \citep{disperse1,disperse2} to reconstruct the cosmic web structure from input galaxy catalogs. 
There are three main steps in a typical {\disperse} workflow for 3D input data (the steps are nearly identical for 2D data).

\vspace{-5pt}

\begin{enumerate}
\itemsep-2pt
    \item Supply point-set inputs -- in our case, catalog of galaxies with ${\logms}\geq8$) -- to the {\it delaunay\_3d} program. In this step, a technique called the Delaunay Tessellation Field Estimator \citep[DTFE; ][]{SV00,VS09} is used to tessellate the entire input volume into tetrahedrons, with the positions of individual galaxies as vertices. The output is a 3D network file, containing the density field estimate {\it at the locations of the vertices (i.e., galaxy positions)}, which is then linearly interpolated for the density field estimate of other spatial locations.
    \item Apply {\it mse}, the main algorithm in {\disperse}, to the 3D density field grid (i.e., the network file containing the density field estimate). Here, the gradients of the density field are computed, and critical points are determined where the gradient is zero, corresponding to the voids (minima), saddle points, and nodes (maxima) of the density field. Filaments are defined as segments connecting maxima to saddle points following ridges of the density field. The output of {\it mse} is a so-called ND skeleton file, which contains the filamentary network described by a list of critical points, segments that make up filaments, and filaments that originate and lead to maxima.
    \item The final step is to apply the ND skeleton output to {\it skelconv} so that the output can be converted to a non-binary format that humans can read. 
\end{enumerate}
\vspace{-5pt}

We reconstruct the cosmic web in each TNG snapshot from $z=0$ to $z=4$ following these steps above. In addition to this nominal procedure, we make several choices to improve the reconstruction. 
In step 1 (during the DTFE density field estimation), following \citet{Malavasi22}, we also smooth the density field value at the position of each galaxy.
We average this with the density field value for all galaxies whose tetrahedrons share an edge with this galaxy.
This is so that we minimize contamination by noise and detection of spurious features on small scales. Similar to \citet{GE24}, we find a non-negligible fraction of unphysically small ($<$0.1 Mpc) output filaments from DTFE density fields that are not smoothed. 
In step 3, we apply smoothing to the position of the segments of the filamentary skeleton.
Here, the critical points are fixed, and the coordinates of each point along a filament are averaged with that of its two neighbors so that sharp and/or unphysical shapes (caused by shot noise) are not seen in filament segments.

Persistence is the key parameter used by {\disperse} to determine the significance of topologically identified structures. {\disperse} identifies topologically significant pairs of critical points, also known as persistence pairs, and associates each pair with a persistence value. This value is defined as the density ratio at the two critical points. The persistence value measures the robustness of the identified topological structure to local fluctuations in the density field due to shot noise in the input data. A higher persistence structure is more significant and robust than a lower one. In practice, the persistence threshold can be set directly in {\it mse} using the option ``-cut'' in terms of a persistence value or using the option ``-nsig'' in terms of the probability that a given structure appears in a random field. In Section~\ref{sec:identify}, we discuss the choice of persistence value when using different methods.

\vspace{-5pt}

\subsection{The MCPM model and Application to TNG}
\label{sec:mcpm}

The MCPM algorithm was first introduced in \citet{Burchett20}. It is an agent-based generative model designed to emulate the growth of the unicellular organism {\it Physarum polycephalum}, commonly known as {\it slime mold}. This organism is known to explore its environment for food sources and shape itself into highly intricate networks to connect them. Scores of researchers across various fields such as computational science, biology, and design engineering, have used this behavior as inspiration for solving a wide range of spatial problems, such as navigating labyrinths \citep{Nakagaki00}, designing transportation networks \citep{Tero10}, and understanding human cognitive patterns and creativity \citep{Adamatzky13}.

The MCPM algorithm is an extension of the model by \citet{Jones10}, who proposed a virtual Physarum machine with a simulated slime mold network.
Similarly to their work, MCPM models the slime mold as discrete particle-like ``agents'' that swarm toward input ``food sources''.
The input data are usually positions of galaxies or halos but can also be other data types such as positions of DM or gas particles in simulations. 
Each input data point deposits markers proportional to the weight of the point, typically stellar or halo mass, which the agents are guided towards. Agents have a higher probability of moving towards regions with higher deposit values. 
The model is run across many time-steps, with the agents moving at each time-step according to the probabilities mentioned above. 
We refer the readers to \citet{Burchett20}, \citet{Elek21}, and \citet{Elek22} for more details.

Although the \citet{Jones10} model was designed so that agents follow the direction of maximum deposit, this would generate extremely condensed and less inter-connected filamentary networks than are observed and expected from theory, therefore not capturing the true complexity of the multi-scale structures of the cosmic web \citep{Burchett20,Elek22}.
Instead, in MCPM, the extracted 3D network is a probability density of the cosmic matter distribution, enabled by three different probability distributions that determine the direction and distance an agent moves relative to its initial state at each time-step of the algorithm. These require the agents to navigate towards large pools of deposits preferentially, connecting the galaxies/halos (via input data deposits) to each other and reinforcing existing pathways (via agent deposits), as well as ensuring that angular momentum is conserved and agents do not end up stuck in local deposit pools. 
\citet{Elek22} showed that MCPM misses less than $\sim0.1\%$ of the input data points, meaning that virtually all of the input galaxies are reliably contained by the reconstruction.

The main output of MCPM is a continuous 3D trace density field. This is the equilibrium density field of the agents, which is created by superimposing all individual agent trajectories across hundreds to thousands of time-steps of the model. 
The trace field can thus be interpreted as a total probability distribution of the agents.
MCPM data fitting is a semi-supervised process aided by a graphical data visualization component, through the visualization software {\sc Polyphorm} \citep{Elek21}.
It allows a user to tune the model's hyperparameters in real-time to obtain the best possible fit, which optimizes a fitness/objective function that determines the level of ``connectedness'' of a fit. Mathematically, this function is a maximum likelihood estimator of the trace density field normalized by each input data point's weight (galaxy mass). 
Since the objective of the MCPM algorithm is to construct an optimal transport network given the input data points, the fitness function ensures that (a) almost all, if not all, data points, are included in the network, and (b) the weight of the data points impact the strength of the connections in the network, i.e., agents have a higher probability of being attracted towards higher mass galaxies.
The MCPM fit is considered to be converged when the fitness function plateaus for a given set of hyperparameters. 
We note that the MCPM fit is fairly robust to small changes in the interactively adjusted hyperparameters \citep[see, e.g.,][]{Burchett20}.

However, maximizing the fitness function does not necessarily guarantee the most physically accurate description of the cosmic density field. A fit with a high fitness function can be, in contrast to a network of interconnected filaments, a series of disconnected islands. The interactive visualization process allows a user to visually verify that such disconnected islands are not produced. 
We recognize that the ideal fitting procedure would rigorously optimize the fitness function against the reference TNG DM density field. While beyond the scope of this work, we aim to address this issue in the future to obtain a precise mathematical formulation of the fitness function and transform MCPM from a semi-supervised to a fully automated reconstruction tool. 

\vspace{-5pt}

\subsubsection{Calibrating Input Parameters}
\label{sec:hyper}

In order for the MCPM trace density field to be a robust proxy of the cosmic matter density, we require a procedure to find the set of input parameters that allow for the best mapping of the trace density at each 3D location to the corresponding physical DM mass density. 
In \citet{Burchett20}, such a mapping between the trace density of MCPM and the DM overdensity was performed for the large Bolshoi-Planck (BP) $N$ body cosmological simulation \citep{Klypin16} at $z\!=\!0$. They found a strong correlation between the trace density and the DM overdensity at higher trace densities, but a roughly flat DM overdensity for very low trace densities. This is due to the much larger dynamic range of the MCPM trace values than the BP DM density values.

Since the TNG100 simulation has different physical prescriptions and volumes compared to BP, 
we calibrate the MCPM hyperparameters against the DM density fields on TNG100. To identify the optimal hyperparameters that result in the best correlation of MCPM trace density to DM density, we run MCPM with a combination of hyperparameter values at each of the snapshots we study. 
We systematically vary the following hyperparameters: 

\vspace{-5pt}
\begin{itemize}
\itemsep-2pt
    \item {\it sampling sharpness}: the geometric definition of the output density field. A lower sharpness value will produce more interconnected and ``fuzzier" filaments, while higher sharpness values will result in cylindrical or tubular filaments that are poorly connected.
    \item {\it persistence coefficient}: the level of robustness of the filamentary structure to shot noise in the input data. This is similar (though not identical) to the persistence parameter implemented in {\disperse} as mentioned above. A lower coefficient leads to sharper structure and a higher coefficient leads to fuzzy, less-defined blobs.
    \item {\it sense distance} and {\it move distance}: these define the distance at which agents can detect the input data and move towards them in each iteration, respectively.
    These are adjusted such that agents do not overshoot the actual density structure.  If these distances are too large, that can cause the agents to trace paths that do not follow the galaxy distribution.
    \item {\it sense angle} and {\it move angle}: these define the angle at which agents can detect data and move towards them at each iteration, respectively. Decreasing these angles results in straighter filamentary networks traced by the agents, increasing them causes filaments to be rounder.
    
\end{itemize}
\vspace{-5pt}

Our approach is to run MCPM with at least 10 different combinations of the above parameters -- visually verifying that the fit converges to a reasonably high fitness function and produces a filamentary structure -- for each of the redshifts $z=0$, 0.5, 1, 2, 3, and 4. We then compare the trace density field against the DM density field using two statistical estimators: the Pearson $r$-value and the root mean squared error (RMSE), with the goal being to minimize the RMSE score and maximize the Pearson $r$-value. We use these experiments to identify the optimal hyperparameters for the fiducial MCPM fits at each redshift.

As a starting point for the hyperparameter values, we use previous knowledge of running the MCPM algorithm as well as {\it a priori} knowledge based on physical considerations. 
From the BP simulation, the optimal {\it sampling sharpness} was found to be 2.5 at $z=0$ and 2.2 at $z=0.5$ \citep{Wilde23}. 
Since the cosmic web becomes more condensed over time, i.e., matter collapses into denser filamentary networks owing to the gravitational pull of structures of ever-increasing masses, we decrease the {\it sampling sharpness} with increasing redshift.
The {\it persistence coefficient} parameter is somewhat dependent on the sparseness of the input data, but previous applications of MCPM \citep[e.g.,][]{Burchett20,Elek22,Wilde23} have found it to be optimally around 0.9.
The {\it sense distance} parameter should roughly follow the volume of the datacube and resolution of the trace field. Based on SDSS observations and BP simulation, \citet{Burchett20} and \citet{Wilde23} show that $\approx$2--2.5 Mpc is a good estimate of the {\it sense distance} at $z\leq0.5$. 
A {\it sense angle} of 20 degrees has been used as a default in most previous applications of MCPM. 
We set the {\it move distance} to be $\sim$5\% of the {\it sense distance} in each case, as larger relative values start to diffuse the MCPM agents to unstructured blobs. For similar considerations, we also fix the {\it move angle} to be half the {\it sense angle}.
We consider the following ranges for our hyperparameters: {\it sampling sharpness} = 1--5, {\it persistence} = 0.8--0.95, {\it sense distance} = 1--4 Mpc, 
and {\it sense angle} = 10--50 degrees.

\begin{figure*}[htbp]
\centering
\gridline{\fig{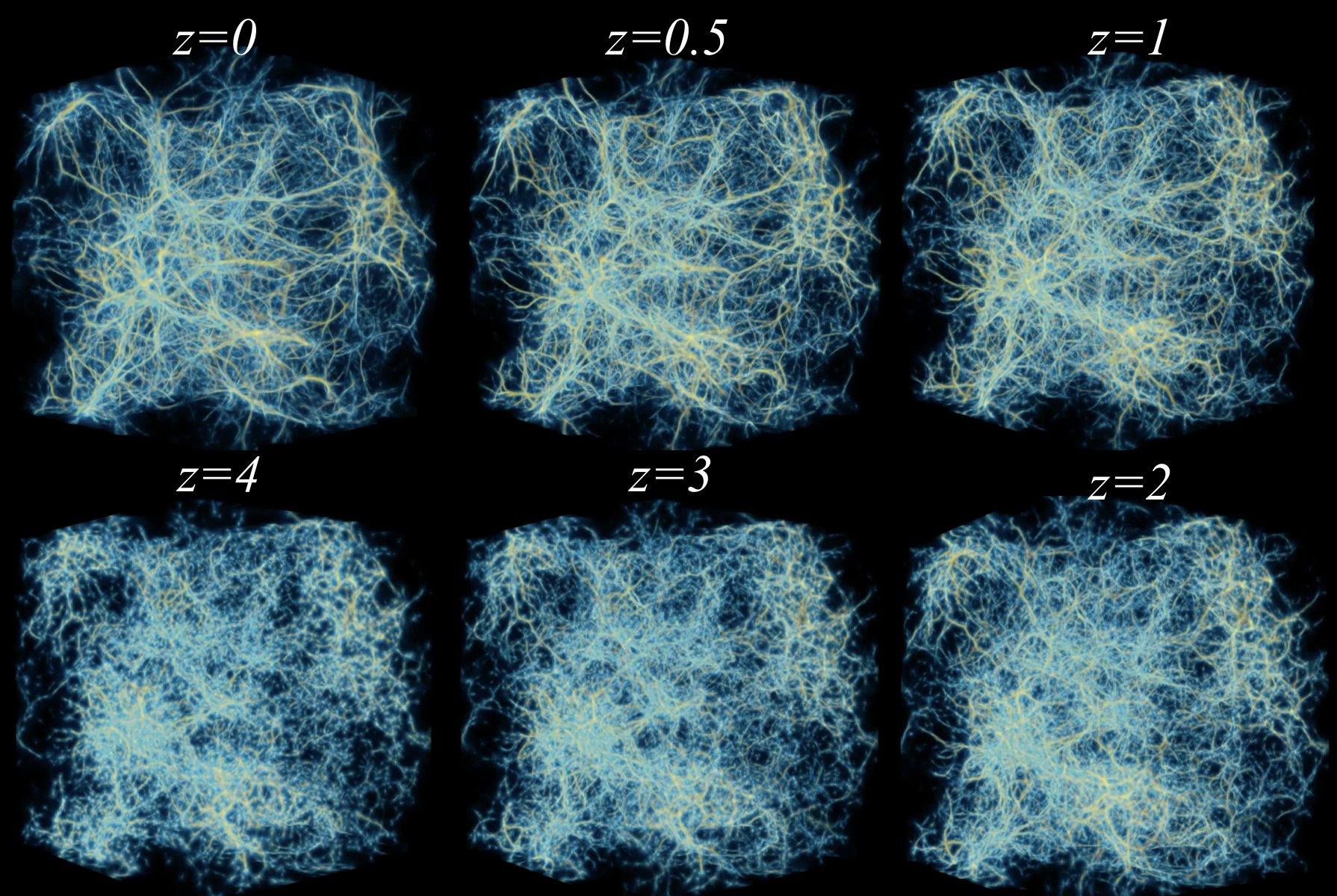}{0.95\textwidth}{}
}
\vspace{-20pt}
\caption{The MCPM trace density field output for TNG100 at redshifts (clockwise from top left) $z=0$, 0.5, 1, 2, 3, and 4. In each cube, 
the density field is represented by the yellow-to-blue colorbar (yellow represents regions of higher density).  
The hyperparameters used for these fits are presented in Table~\ref{tab:mcpmparams}.
Figures were rendered with the {\sc Polyphorm} visualization tool for MCPM \citep{Elek21}.}
\label{fig:mcpmtrace}
\end{figure*}

Using the {\sc Polyphorm} software, we construct the MCPM density field at a given redshift by applying the locations of galaxies with ${\logms}\geq8$ as inputs on a grid with $512^3$ voxel resolution, which corresponds to a $\approx$0.25 Mpc physical scale. This resolution has been commonly used for volumes on the order of $\approx$100 Mpc and \citet{Elek22} showed that increasing the voxel resolution to $768^3$ or $1024^3$ decreases Monte Carlo noise but does not significantly alter the output density field.  For each of our runs, we choose 5 million agents in the fitting process, resulting in a resolution of $\lesssim$30 
voxels per agent on average. Choosing the number density of agents within a factor of 2--3 does not noticeably affect the output density field.
Additionally, 
\citet{Elek22} demonstrated that cosmic density estimation with MCPM is highly robust to changes in the number of inputs -- even if by an order of magnitude.
The weight of each galaxy is proportional to its stellar mass.
We use our group's custom-made {\sc pyslime} software\footnote{\url{https://github.com/jnburchett/pyslime}} to read in the MCPM density field outputs for each redshift.

We apply some additional post-processing to the output trace density field from MCPM. We first smooth the density fields on different resolutions -- corresponding to different spatial scales -- using a Gaussian kernel, with the standard deviation defined by the smoothing scale. 
We smooth on scales corresponding to the $512^3$, $256^3$, $128^3$, and $64^3$ voxel resolutions. 
For scales larger than the original $512^3$ resolution, we downsample the density field by a factor of 2 (for a $256^3$ resolution grid), 4 ($128^3$), and 8 ($64^3$). 
We calculate the MCPM overdensity at each spatial location by dividing the trace density at that location by the mean of the entire volume. Lastly, we repeat this post-processing for DM density fields and compare the RMSE scores and Pearson $r$-values.

We find that smoothing the MCPM density fields reduces Monte Carlo noise and that downsampling the fields results in better correlations with smoothed and downsampled DM density fields on the same scales. 
Going from smoothing scales of $512^3$ to $256^3$, the RMSE scores for a given $z$ decreased by $\sim$0.2--0.3 and the $r$-values increased by $\sim$0.15--0.2 on average. Going from smoothing scales of $256^3$ to $128^3$, the corresponding decrease in RMSE scores was $\approx$0.05 and the corresponding increase in $r$-values was $\approx$0.05--0.1. 
However, too much smoothing and downsampling blurs over real features of the cosmic web and reduces the level of detail of the density field. 
After visual analysis, we determined that the $256^3$ resolution scale was the best compromise for the scale where the MCPM density field both correlates well with the DM field and is high enough in resolution to identify the less prominent large-scale filamentary structures. Regardless, our conclusions about the statistics of identified filaments and their impact on galaxy formation processes does not change when we adopt a $128^3$ resolution scale.

We measure the correlation between DM overdensity and MCPM overdensity for different hyperparameters on $256^3$ grids at each redshift. While a handful of combinations of hyperparameters cause MCPM overdensity to correlate less well with DM overdensity, the vast majority of combinations generated well-correlated overdensity values.
We confirm visually and quantitatively that the vast majority of hyperparameter values in MCPM lead to a good estimation of the cosmic density field and to the MCPM density being a robust proxy of the underlying cosmic matter distribution.

We show in Table~\ref{tab:mcpmparams} the results of our experiment at each redshift -- the optimal hyperparameters that maximized the Pearson $r$-value and minimized the RMSE score.
The best RMSE scores ranged from 0.53 to 0.7, and the best Pearson $r$-values ranged from 0.49 to 0.58.
The {\it sampling sharpness} and {\it sense distance} parameters affect the output density field most strongly, while the {\it sense angle} affect it the least. 
As a sanity check, we verify that there is a smooth progression in parameter values as we go from earlier to later times (as opposed to sudden discontinuities at a given redshift). The optimal {\it sampling sharpness} parameter increases from 1.5 at $z=4$ to  2.5 at $z=0$, and the {\it persistence coefficient} increases from 0.825 to 0.86 between $z=4$ and $z=0$, both of which reflect matter condensing into well-defined structures more with time \citep[e.g.,][]{Cautun14,Martizzi19}.
The optimal {\it sense distance} increases from 1.5 Mpc to 2.5 Mpc at $z=4\rightarrow1$, which is a consequence of cosmic structures growing larger with the Hubble flow as more time passes. The {\it sense angle} was the only parameter whose optimal value was always the same (20 degrees) at each redshift. Recall that we fix the {\it move angle} at 50\% of the {\it sense angle} and the {\it move distance} at 5\% of the {\it sense distance}.

We present the output MCPM trace density fields constructed from these parameters in Figure~\ref{fig:mcpmtrace}.
The color-coding yellow-to-blue represents the trace density value from high to low. 
Clearly, the cosmic density fields at higher redshifts appear more fragmented and less condensed than those at lower redshifts.
We visually compared the MCPM density fields from different combinations of hyperparameters and found that the vast majority of them look similar to the fiducial case with our determined optimal parameters (Table~\ref{tab:mcpmparams}).
Despite some of these density fields diverging significantly from the optimal case in a visual sense, they all statistically correlated better than the DTFE density field with the DM density field.

\begin{deluxetable}{cccccc}[hbtp]
\centering
\tablewidth{0.48\textwidth}
\tabletypesize{\small}
\tablecaption{
Hyperparameters of the MCPM density field computations at each redshift
}
\label{tab:mcpmparams}
\tablehead{
\colhead{$z$} & 
\colhead{sampling} &
\colhead{persistence} &
\colhead{sense\tablenotemark{\scriptsize a}} &
\colhead{sense\tablenotemark{\scriptsize a}} \\[-5pt]
\colhead{} & 
\colhead{sharpness} &
\colhead{coefficient} &
\colhead{angle [deg]} &
\colhead{distance [Mpc]}
}
\startdata
0 & 2.5 & 0.86 & 20 & 2.5 \\
0.5 & 2.25 & 0.85 & 20 & 2.0 \\
1 & 2.0 & 0.84 & 20 & 1.8 \\
2 & 1.8 & 0.835 & 20 & 1.5 \\
3 & 1.65 & 0.83 & 20 & 1.5 \\
4 & 1.5 & 0.825 & 20 & 1.5 \\
\tableline
\enddata
\tablenotetext{\scriptsize a}{We always fix the {\it move angle} at 50\% of the {\it sense angle} and the {\it move distance} at 5\% of the {\it sense distance} (see text).}
\end{deluxetable}
\vspace{-40pt}

\vspace{-5pt}

\subsubsection{Comparing Density Fields}
\label{sec:denscomp}

\begin{figure*}[htbp]
\vspace{-10pt}
\centering
\gridline{\fig{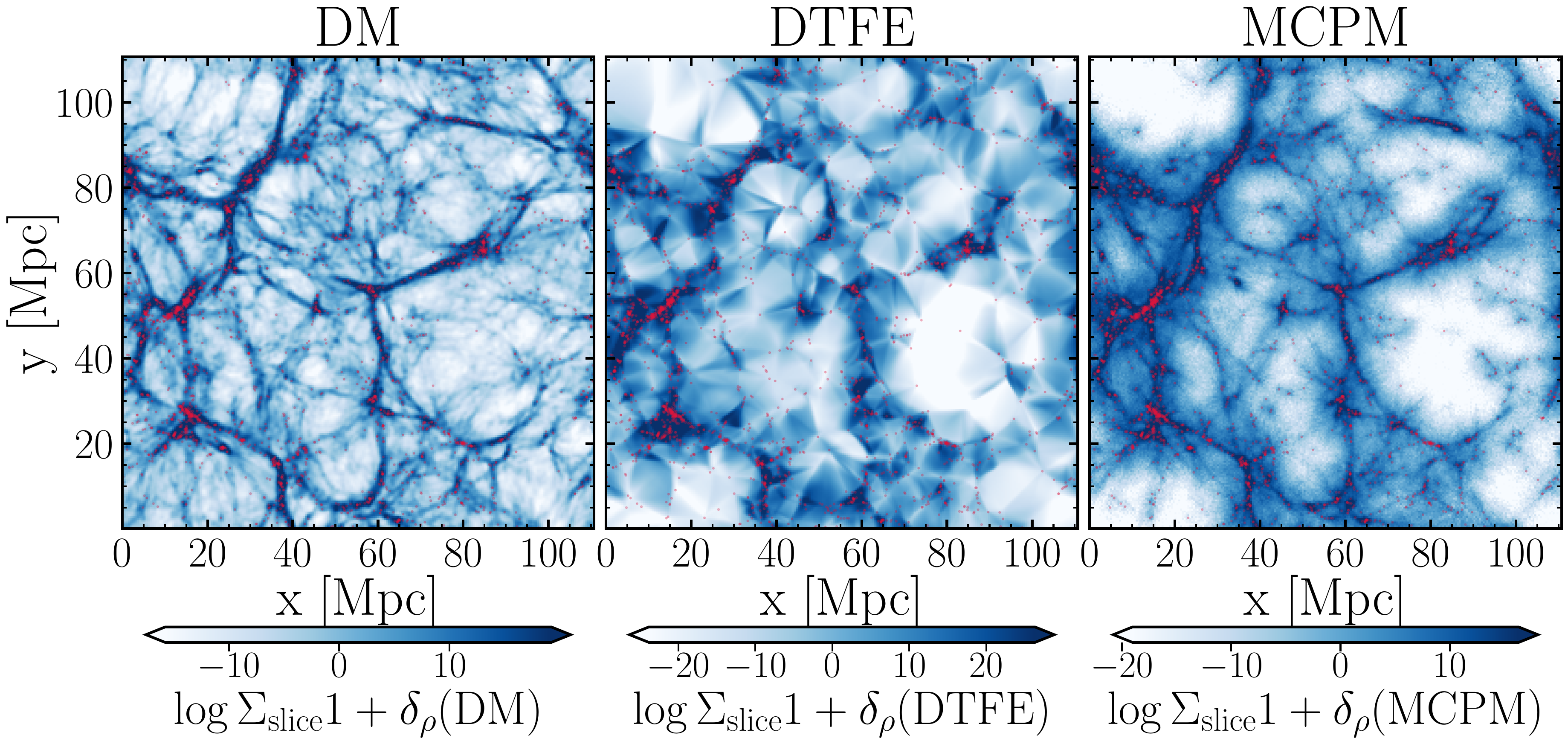}{0.98\textwidth}{}
}
\vspace{-25pt}
\gridline{\fig{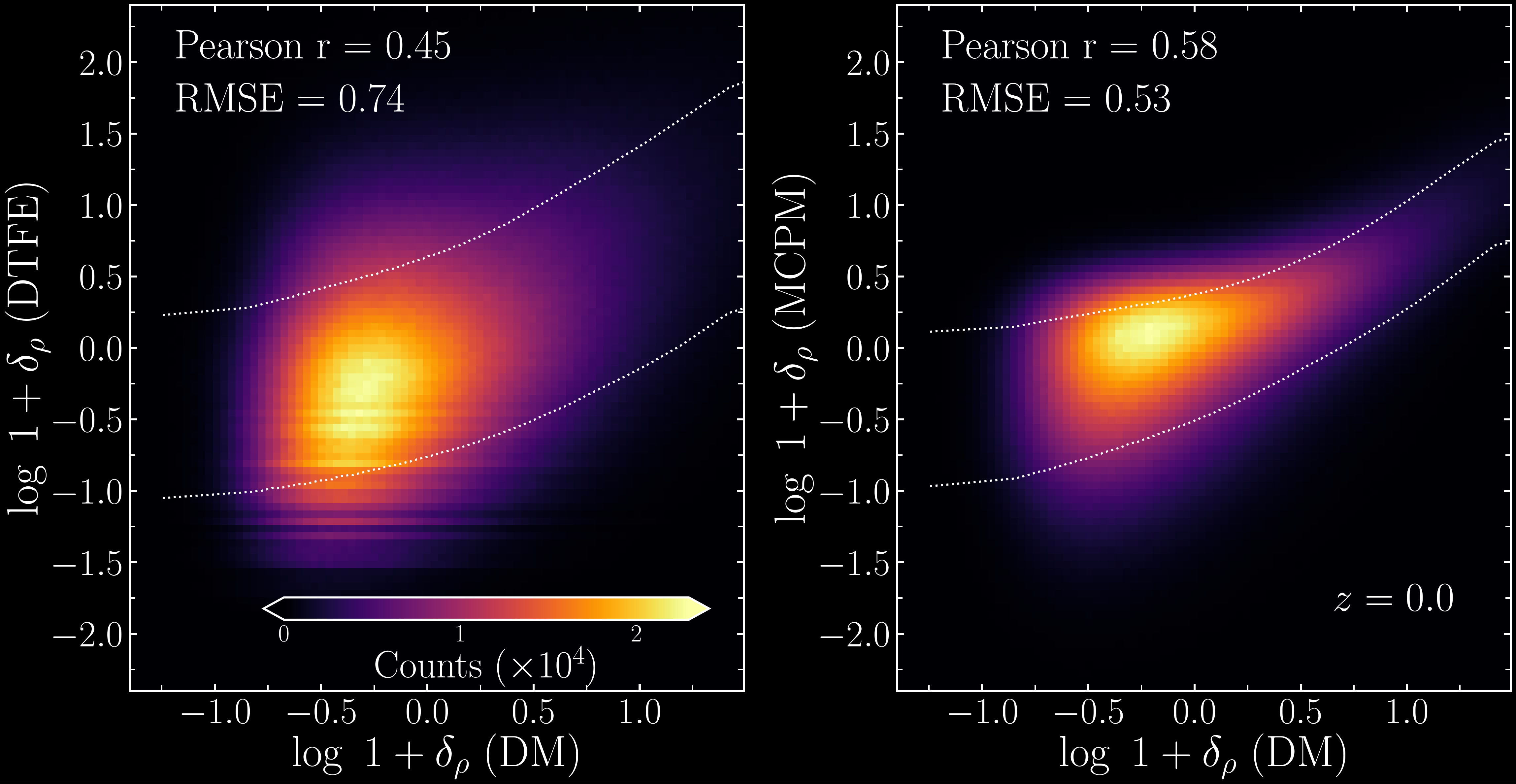}
{0.7\textwidth}{}
}
\vspace{-25pt}
\caption{Comparison between the continuous DM, DTFE, and MCPM density fields at $z=0$ in TNG.
{\it Top:} Visualization of the DM (left), DTFE (middle), and MCPM (right) density field projected in the x-y plane. The blue colorbar represents the overdensity for a given density field along a 10 Mpc thick slice, and red circles represent the positions of galaxies (sized in proportion to their stellar mass). 
{\it Bottom}: 2D histograms showing the mapping between DM and DTFE (left) and MCPM (right) overdensity. The dotted curves show the running middle 68\% of distributions in DTFE or MCPM overdensity along DM overdensity bins.
The Pearson $r$-value and RMSE score for each correlation are indicated in the top left of each panel. 
Both visually and statistically, the MCPM density field is a superior to the DTFE density field at approximating the underlying DM distribution.
}
\label{fig:densitycomp}
\end{figure*}

MCPM produces a continuous estimate of the 3D density field at each pixel of the reconstructed volume. The density estimate at $>\!10^8$ spatial locations of the TNG100 volume (for a full $512^3$ grid) is a significantly richer description of the cosmic density field compared to the density estimate at the $\sim20,000-50,000$ locations of individual galaxies, and linearly interpolated estimates elsewhere, from the DTFE density estimator.
This is despite using the same input galaxy catalogs in both density estimators. 
The density field produced by MCPM, therefore,
allows for a more accurate characterization of density far from galaxies to voids, as it does not perform any interpolations of the density field in locations beyond galaxies. Thus, the MCPM density field is a natural candidate for the input density field in {\disperse} and, therefore, begs a comparison with the DTFE density field.

We make a quantitative comparison to test the fidelity of both the MCPM and DTFE density fields constructed from the same inputs and post-processed in the same manner.
To generate a continuous DTFE density field from the distribution of galaxies at each redshift, we make use of The Delaunay Tessellation Field Estimator public software (hereafter {\sc dtfe} package)\footnote{\url{https://github.com/MariusCautun/DTFE}}, as described in \citet{Cautun19}, and based on the method developed by \citet{SV00}.
The {\sc dtfe} package implements a mass-normalized Delaunay Tessellation on a set of discrete inputs to construct a continuous, interpolated mass density field. This implementation of DTFE is similar to the {\disperse} implementation, except that it produces a continuous density field grid (via interpolation) instead of a discrete set of density values at the locations of the inputs.
Just as for the MCPM density estimation, we apply the locations of ${\logms}\geq8$ galaxies at a given redshift as inputs to the {\sc dtfe} package, with the weight of the galaxy proportional to its stellar mass. We construct the DTFE density field on a $512^3$ grid using periodic boundary conditions and calculate the volume averaged density inside a grid cell using Monte Carlo integration, taking 1000 random sample points per grid cell.
We find that using non-volume average density fields introduces significant shot noise, especially in high density regions, and that varying the number of sampling  points does not have a significant effect on the output density field.

We then smooth and downsample the DTFE density fields to resolutions of $256^3$, $128^3$, and $64^3$, as we did for the MCPM density fields, and compute the overdensity at each spatial location. We statistically and visually compare the output density field to the DM density field. Similar to MCPM, the RMSE decreases and Pearson $r$-value increases with decreasing grid resolution, but regardless, the correlation between DTFE and DM overdensity is always weaker than that between MCPM and DM overdensity.

In Figure~\ref{fig:densitycomp}, we present a visual (top panels) and statistical (bottom panels) comparison of the DM, DTFE, and MCPM density fields at $z=0$. In the top panels, the same 10 Mpc thick slice is projected onto the x-y plane, and galaxies are represented as red circles (sizes proportional to stellar mass), while the DM overdensity (left), DTFE overdensity (middle), and MCPM overdensity (right) along the slice, smoothed on a $256^3$ resolution scale, are represented by the blue colorbars. 
It is readily apparent that the MCPM density field is a good proxy for the DM density field, while the DTFE density field only accurately describes the higher density regions (consisting of a high concentration of massive galaxies) and poorly describes the density in the lower density regions.
The tessellation of space into tetrahedrons in DTFE washes over the finer details of the true density distribution in much of the volume.

The bottom panels of Figure~\ref{fig:densitycomp} depict 2D histograms of the mapping between DM and DTFE overdensity (left) and MCPM overdensity (right). The dotted curves represent the running middle 68\% (\nth{16}--\nth{84} percentile) of a distribution along DM overdensity. Not only is the Pearson $r$-value higher and RMSE score lower for the MCPM-DM correlation, but the $\pm1\sigma$ spread in MCPM overdensity {\deltamcpm} for a fixed DM overdensity {\deltadm} is much smaller ($\approx$0.5 dex on average) than the spread in DTFE overdensity {\deltadtfe} for a fixed {\deltadm} ($\approx$1 dex on average). 
The MCPM-DM relationship is much more linear than the DTFE-DM relationship and MCPM 
more faithfully captures low-density regions than DTFE.
The DTFE overdensities exhibit greater extremes and lower median values than the DM overdensities, while the median, while the extreme range of DM and MCPM overdensity values are quite close to one-to-one.
We find very similar qualitative results at higher redshifts (not shown), leading us to the conclusion that the MCPM method is superior at approximating the true underlying mass distribution in TNG than the DTFE method.

\vspace{-5pt}

\subsection{Identification of Cosmic Web Structures}
\label{sec:identify}

\begin{figure*}[htbp]
\centering
\gridline{\fig{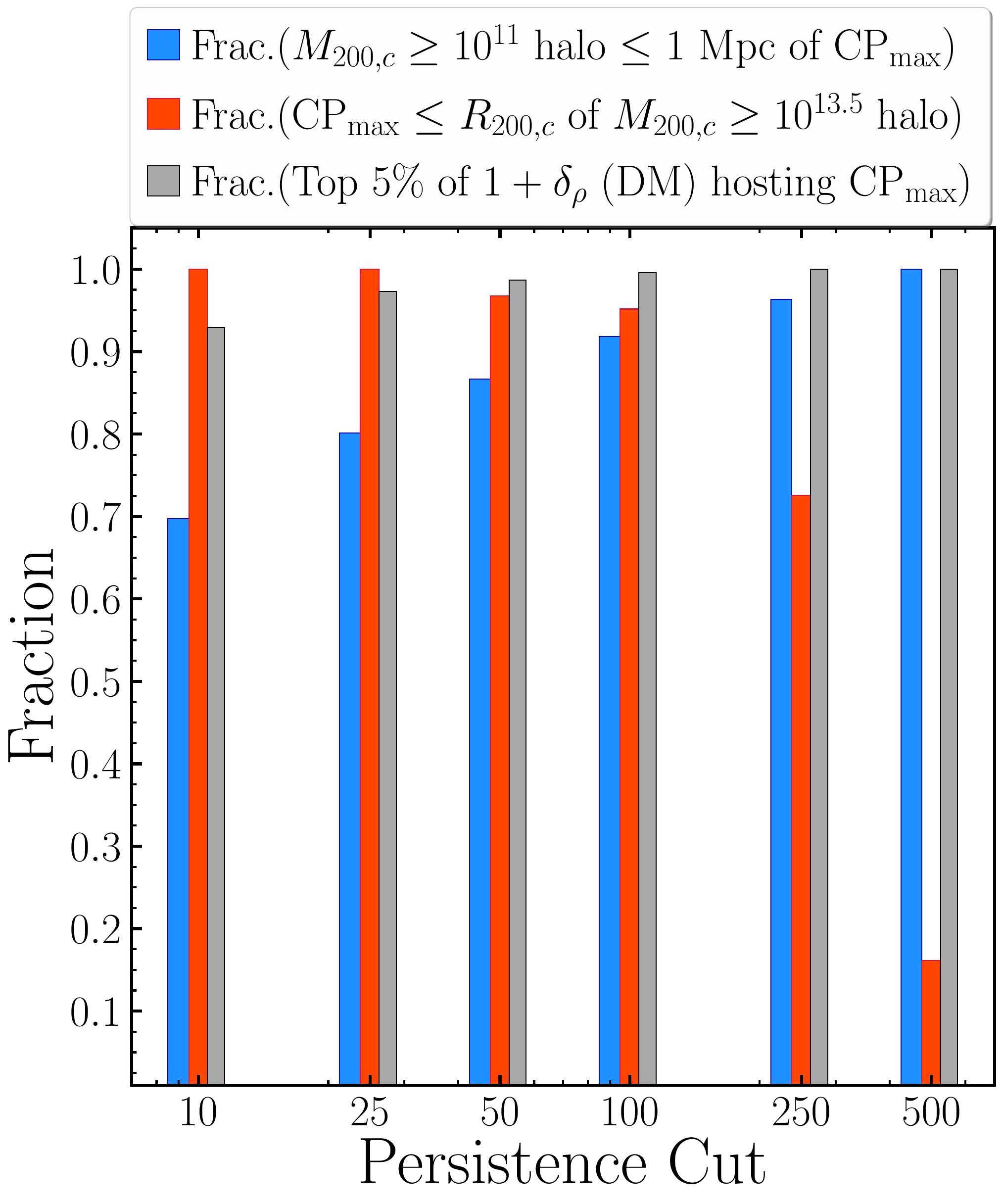}{0.4\textwidth}{}
\fig{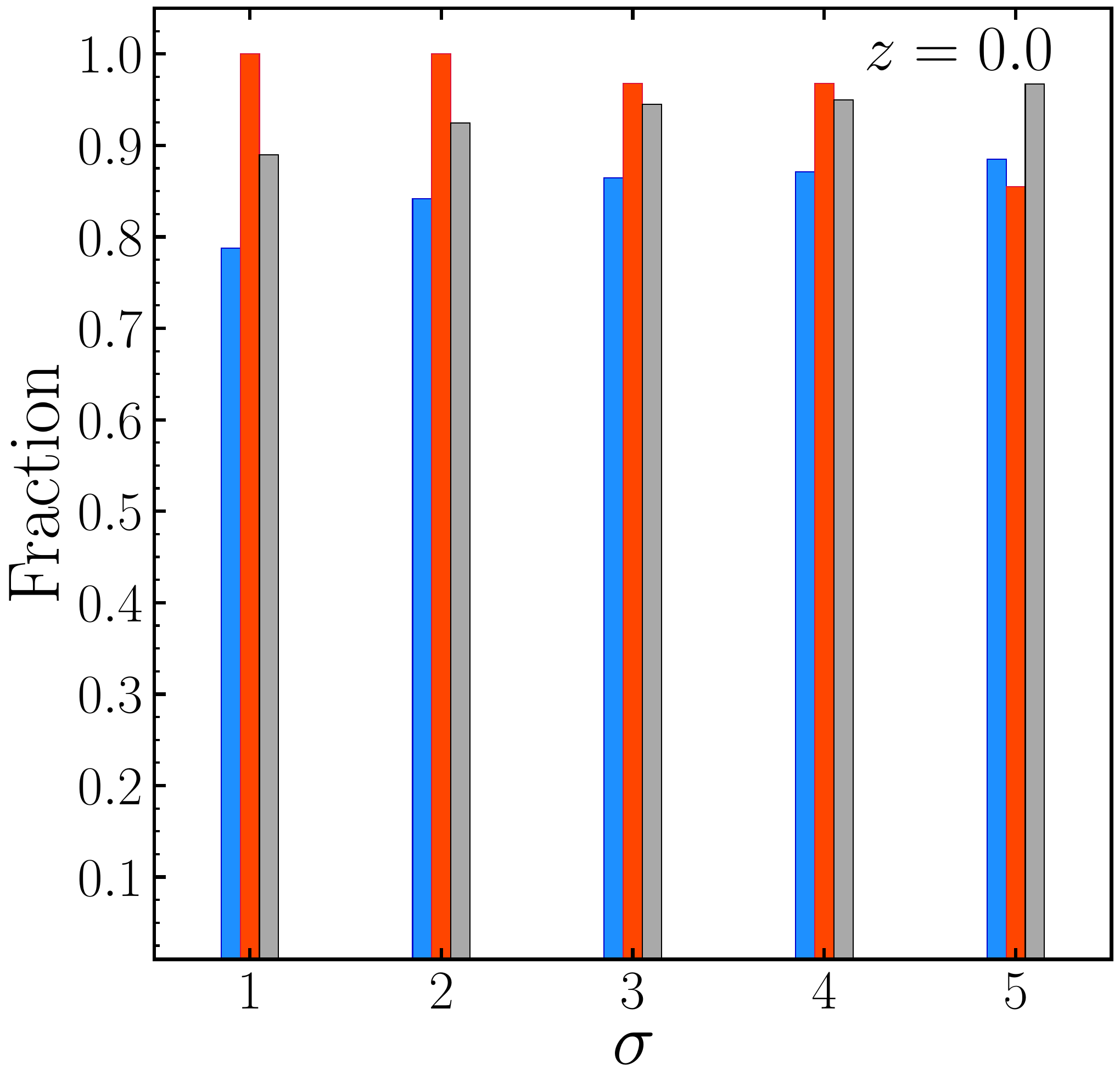}{0.4\textwidth}{}}
\vspace{-25pt}
\gridline{\fig{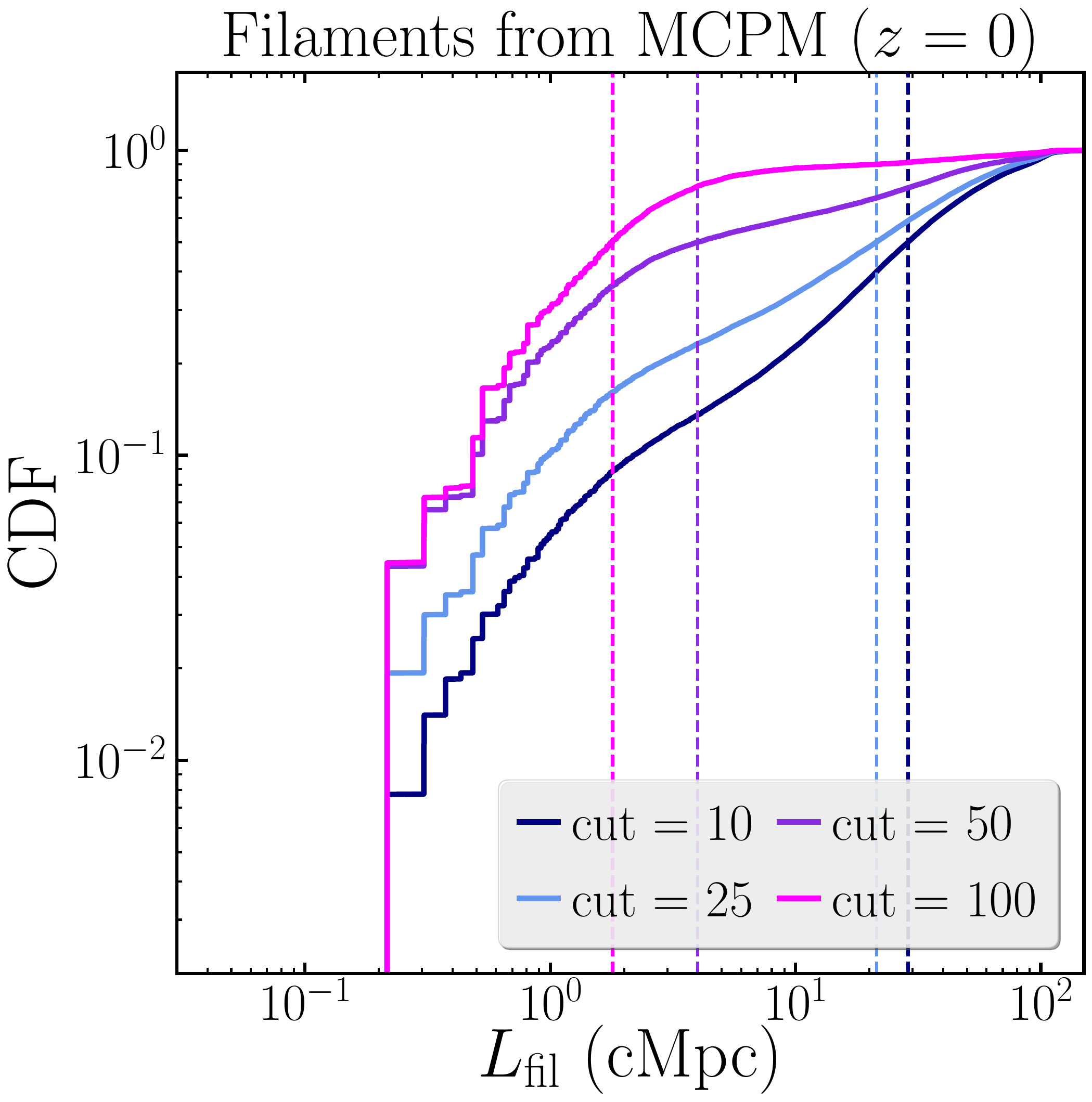}{0.38\textwidth}{}
\fig{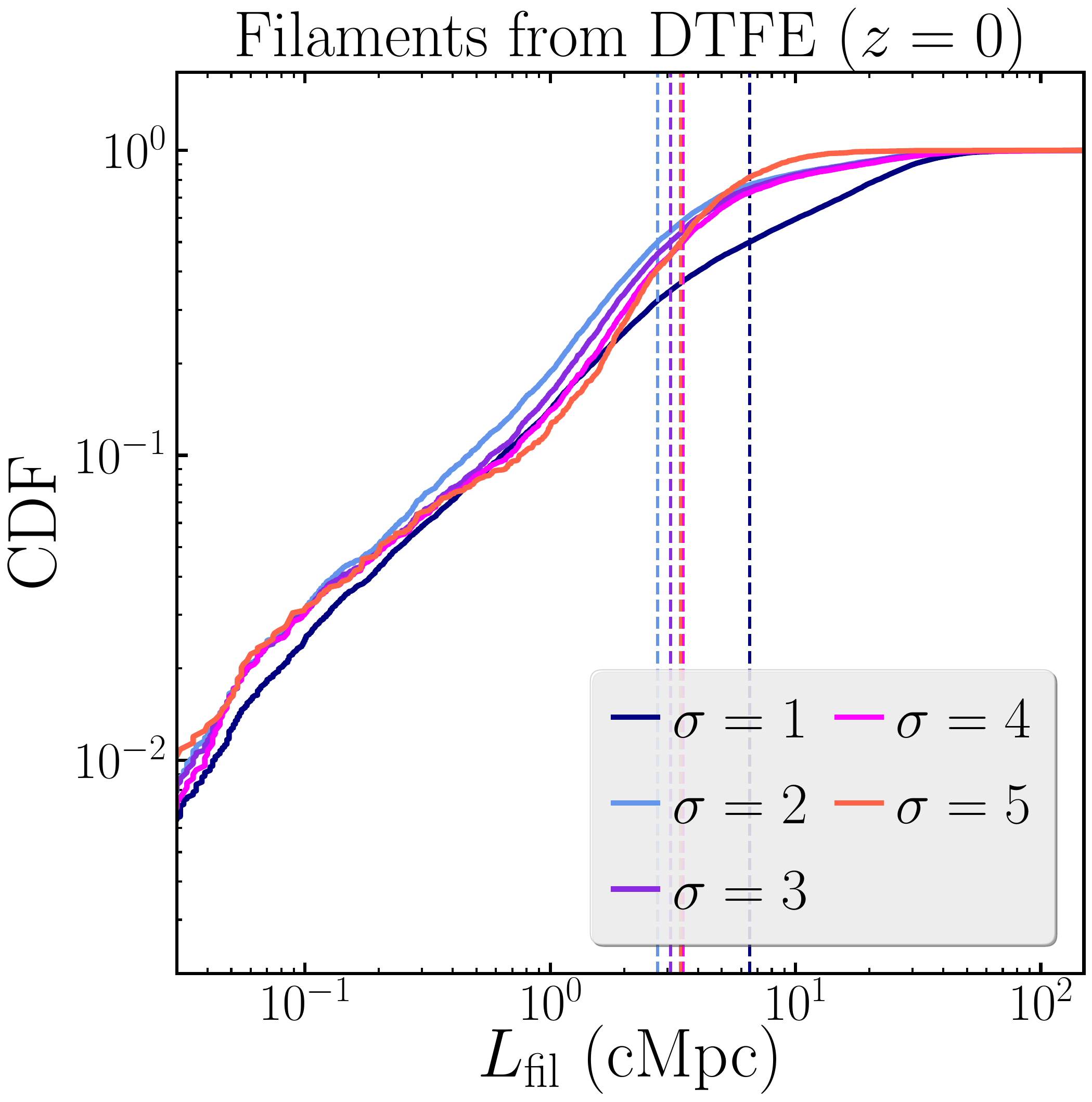}{0.38\textwidth}{}}
\vspace{-25pt}
\caption{
Results of experiments to determine the optimal persistence threshold to identify filamentary structure in {\disperse} at $z\!=\!0$, using the MCPM density field (left) and the DTFE density field (right). 
{\it Top:}
The red bars represent the fraction of ${\logmh}\!\geq\!13.5$ halos containing a node ({\CPmax}) within {\Rh}, the blue bars represent the fraction of ${\logmh}\!\geq\!11$ halos within 1 Mpc of a {\CPmax}, and the grey bars denote the fraction of the highest 5\% of DM overdensities co-spatial with a {\CPmax}.
{\it Bottom:}
Cumulative distribution of the total comoving length of filaments {\Lfil} from MCPM and DTFE using different persistence thresholds. The vertical dashed lines represent the median {\Lfil} for a given persistence. 
We choose persistence cut=25 and $\sigma$=2 as the fiducial persistence level of MCPM- and  DTFE-based reconstruction, respectively (see text for details).
}
\label{fig:cal}
\end{figure*}

Using our downsampled and smoothed MCPM density fields on $256^3$ resolution grids, we undertake the task of identifying cosmic web structures at different snapshots of TNG using the {\it mse} algorithm within the {\disperse}  framework. 
Specifically, in step 2 of the {\disperse} workflow outlined in Section~\ref{sec:disperse}, we apply the continuous MCPM density field instead of the discrete density field estimate from DTFE (prepackaged within {\disperse}).
Next, we convert this grid to the native ND field format of {\disperse} using the {\it fieldconv} program. Finally, we execute the {\it mse} program to identify critical points and the filamentary skeleton of the density field.
We compare the reconstructed filamentary structure from this method to that reconstructed by the traditional method of the discrete DTFE density field.
We note that we attempted to reconstruct filamentary structure using {\it continuous} DTFE density fields smoothed and downsampled on $256^3$ resolution grids as we constructed in the previous section. However, this did not produce physically plausible filamentary structures based on our visual inspection and statistical characterization of output filament lengths. 
We consider additional investigations into the identification of filaments using continuous DTFE density fields to be beyond the scope of this paper; nevertheless, we can conclude that the much-improved filament identifications we report here are not due to the continuous vs. discrete nature of MCPM vs. DTFE.

The identification of discrete structures in {\disperse} depends on the chosen persistence threshold which determines the robustness of topological structures.
However, assessing the accuracy and robustness of these structures presents a challenging problem, as any ideal reconstruction of the cosmic web must satisfy two criteria: (1) {\it completeness} to maximize identifying the complete filamentary structure, and (2) {\it purity} to ensure that the identified filamentary structure is physically sound. To increase completeness, the persistence threshold can be reduced, resulting in the identification of more filaments. However, this may lead to the identification of spurious critical points and filaments not located near any galaxies, which trace enhancements in the matter density field. The balance between completeness and purity of filamentary skeletons is discussed in detail, with numerous examples, in \citet{disperse1}.

In a recent study of the MilleniumTNG simulation \citep{Pakmor23}, \citet{GE24} presented a blueprint to calibrate the persistence threshold by physically motivated priors that address both the completeness and purity of the output skeleton. They found that optimal persistence maximized the fraction of massive groups/clusters that were near a maximum or node of the density field and the fraction of maxima that were near high-mass halos. 
We follow their approach and develop a physically motivated calibration of the optimal persistence threshold by varying the threshold and calculating certain quantities for each threshold, which include (1) the fraction of halos with ${\logmh}\geq13.5$ that contain a node (hereafter {\CPmax}) within their {\Rh}, (2) the fraction of halos with ${\logmh}\geq11$ within 1 Mpc of a {\CPmax}, and (3) the fraction of the highest 5\% of {\deltadm} whose spatial locations contain a {\CPmax}.
For quantity (1), we opted to use a halo mass threshold of ${\logmh}=13.5$ instead of the higher ${\logmh}\simeq14$ threshold adopted by \citet{GE24} due to the very low number statistics of ${\logmh}\geq14$ mass halos in TNG100. 
Approximately 0.5\% of the halos at $z=0$ have ${\logmh}\geq13.5$. 
For quantity (2), we choose a maximum separation of 1 Mpc between {\CPmax} and massive halos as this corresponds to $\approx$2 times the physical smoothing scale of our density fields; this would therefore account for the smoothing of features on small scales in post-processing of the density fields. Individual filament segments are at most 1 Mpc in length, which corresponds to 2 voxels of effective size $\approx$0.5 Mpc.
For quantity (3), we rely on the physical prior that nodes correspond to the most overdense regions of the cosmic matter distribution. While we assess only the top 5\% of the DM overdensity, changing this threshold to the top 2\% or 1\% does not qualitatively change our results below.

To determine the optimal persistence threshold, we aim to maximize the three quantities mentioned above. We also analyze the distribution of comoving filament lengths, denoted by {\Lfil}, by calculating the length of each segment that makes up a filament for different persistence thresholds. This helps us identify which reconstructions generate shorter or longer filaments. 
For reconstructions based on discrete DTFE density fields, we set the persistence level for {\it mse} using the ``-nsig'' parameter, which indicates the minimum number of sigmas above the random field for a structure to be identified. This is the method used by the large majority of {\disperse} users. For reconstructions that use the MCPM density field, we use the ``-cut'' parameter, which is a numerical value of persistence.
For the MCPM density field runs, we test cuts of 10, 25, 50, 100, 250, and for the DTFE density field runs, we test $\sigma=$ 1, 2, 3, 4, and 5.

While it is not possible to simultaneously maximize completeness and purity,
it is important to ensure that we match a large percentage of approximate physical locations between {\CPmax} and massive DM halos and, similarly, between {\CPmax} and peaks in the DM mass distribution. 
The key is to ground our approach in physically motivated reasoning.
In the end, we find that the qualitative conclusions about filament length distributions and the impact of filaments on galactic properties do not change between the best candidates for persistence threshold for either MCPM or DTFE runs.

Figure~\ref{fig:cal} displays the results of our experiments conducted to determine the optimal persistence levels at $z=0$. The top panels contain bar charts to represent fractions (1) through (3) for different persistence cuts for the MCPM density field (left) and different values of -nsig for the DTFE density field (right). 
The bottom panels show cumulative distribution functions (CDF) for selected persistence cuts using the two different methods. Vertical dashed lines represent the median {\Lfil} for a given persistence cut.
We limited the number of cuts displayed in the left panel (for MCPM) to four for increased clarity.

The bar charts indicate that a higher persistence level generally leads to a higher fraction of ${\logmh}\geq11$ halos within 1~Mpc of a {\CPmax}, a higher fraction of matched locations of the highest DM overdensities with {\CPmax},
and a somewhat lower fraction of large group/cluster-mass halos hosting a {\CPmax} within {\Rh}.
For the MCPM density field runs, persistence cuts of 10, 25, 50, and 100 all result in $>$95\% of the {\CPmax} being found within {\Rh} of group and cluster-mass halos. The same is true for $\sigma=1$, 2, 3, and 4 in the DTFE density field runs. 
In the MCPM runs, cuts = 25, 50, and 100 also all generate {\CPmax} that are almost all in the most overdense regions; in the DTFE runs, $\sigma$ = 2, 3, and 4 generate {\CPmax} of which at least 90\% are co-spatial with the most overdense regions. 
All of these parameter values also result in at least 80\% of {\CPmax} being located within 1 Mpc of a ${\logmh}\geq11$ halo.

The filament length distributions give us further insight into which persistence thresholds generate cosmic structures of high completeness and purity.  
For the MCPM runs, we notice that higher persistence cuts of 50 and 100 result in $\sim$25-30\% of the filaments being smaller than 1 Mpc in length. This is a rather large fraction of filaments being only $\sim$2 smoothing lengths across, and many of these could be spurious or small-scale structures that are unresolved. The purpose of this work is to instead identify large-scale structure filaments.
In contrast, cuts of 10 and 25 result in only $\sim$5-10\% of filaments with ${\Lfil}<1$ Mpc. 
For the DTFE runs, there is a much smaller difference between the different {\Lfil} distributions, but all have $\sim$10-20\% of filaments with ${\Lfil}<1$ Mpc. 
We verify that these statistics are not affected by smoothing and downsampled resolution of the density field.

To conclude our experiments, we select cut = 25 and $\sigma$ = 2 for the MCPM and DTFE density field runs, respectively, as the best choice of persistence threshold for completeness and purity. For MCPM, cut = 25 identifies (1) 100\% of {\CPmax} within {\Rh} of ${\logmh}\geq13.5$ halos, (2) 97\% of {\CPmax} at the 5\% most overdense regions of the cosmic matter distribution, (3) $\sim$80\% of {\CPmax} within 1 Mpc of ${\logmh}\geq11$ halos, and (4) only $\sim$10\% of filaments with length ${\Lfil}<$1 Mpc. 
Cut = 50 identifies about 96\%, 86\% and 98\% of {\CPmax} within {\Rh} of ${\logmh}\geq13.5$ halos, within 1 Mpc of ${\logmh}\geq11$ halos, and at the 5\% highest DM overdensity regions, respectively, but this cut results in $\sim$25\% of the filaments being less than 1 Mpc in length.
Similarly for DTFE, $\sigma$ = 2 identifies (1) 100\% of  {\CPmax} within {\Rh} of ${\logmh}\geq13.5$ halos, (2) $\gtrsim$90\% of {\CPmax} at the locations of the 5\% highest DM overdensities, (3) $\gtrsim$80\% of {\CPmax} within 1 Mpc of ${\logmh}\geq11$ halos, and (4) $\sim$10\% of filaments with length ${\Lfil}<$1 Mpc. 
$\sigma$ = 3 satisfies most of these criteria by similarly high fractions, albeit $\approx$5\% of the {\CPmax} are not within the virial radius of ${\logmh}\geq13.5$ halos. 
Our fiducial choice of $\sigma=2$ is the same as that of \citet{GE24} for their MilleniumTNG reconstruction. However, it differs from the cut $\sigma=3$ that we used in \citetalias{Hasan23} based on previous work on TNG100 \citep[e.g.,][]{Malavasi22}. We emphasize that the subsequent analysis we perform in this paper produces slightly different quantitative results, but not qualitative results, between choosing $\sigma$ = 2 or 3 in the DTFE runs, and analogously, between cut = 25 or 50 in the MCPM runs.
Finally, we make the crucial observation that the MCPM density field runs identify an order of magnitude more filaments than the DTFE density field runs while maintaining accurate reconstruction of the skeleton through physical matching between {\CPmax} and massive halos. This will have a marked effect on our results as we elaborate in more detail below.

We apply the same persistence cuts for the MCPM and DTFE runs for all the higher redshift cosmic web reconstructions as well, since we aim to trace structures on the same scales and cosmic density contrasts as those at $z=0$.
Although we only present the results of the physical calibration procedure at $z=0$, we also perform such an experiment at all the higher redshift snapshots. 
Similar to $z=0$, the identified structures for cut = 25 and $\sigma$ = 2 are physically robust at higher $z$, including most of the $\sim$0.5\% most massive halos at a given $z$ hosting {\CPmax} in their virial spheres and most of the {\CPmax} being co-spatial with the 5\% most overdense regions at a given $z$. However, the matched fractions of massive halos and {\CPmax} and DM overdensities and {\CPmax} do decrease with increasing redshift for any given persistence threshold. This is because the mean density of the universe and the matter density contrast decreases with increasing redshift, leading to more homogeneous density fields and consequently smaller density peaks that are less likely to be identified as nodes of the cosmic web.

We also experiment with the smoothing parameter in the {\it skelconv} program, which effectively averages the locations of the individual points that make up a filament segment. 
For our MCPM and DTFE density field runs, we applied a smoothing of 1, 2, and none to the filamentary skeleton (smoothing higher than 2 levels washes out local geometrical structure of filaments). We find little to no impact on the visual structure of the reconstructed filaments, and both the qualitative and quantitative results of the statistics of the filaments, such as the {\Lfil} distribution,
agree with \citetalias{Hasan23}.
We have selected a smoothing of 1 as our fiducial value for both the DTFE and MCPM runs, as this reduces unphysically sharp filament segments.

\vspace{-5pt}

\subsection{MCPM vs. DTFE Filaments}
\label{sec:dtfevmcpm}

\begin{figure*}[htbp]
\vspace{-10pt}
\centering
\gridline{\fig{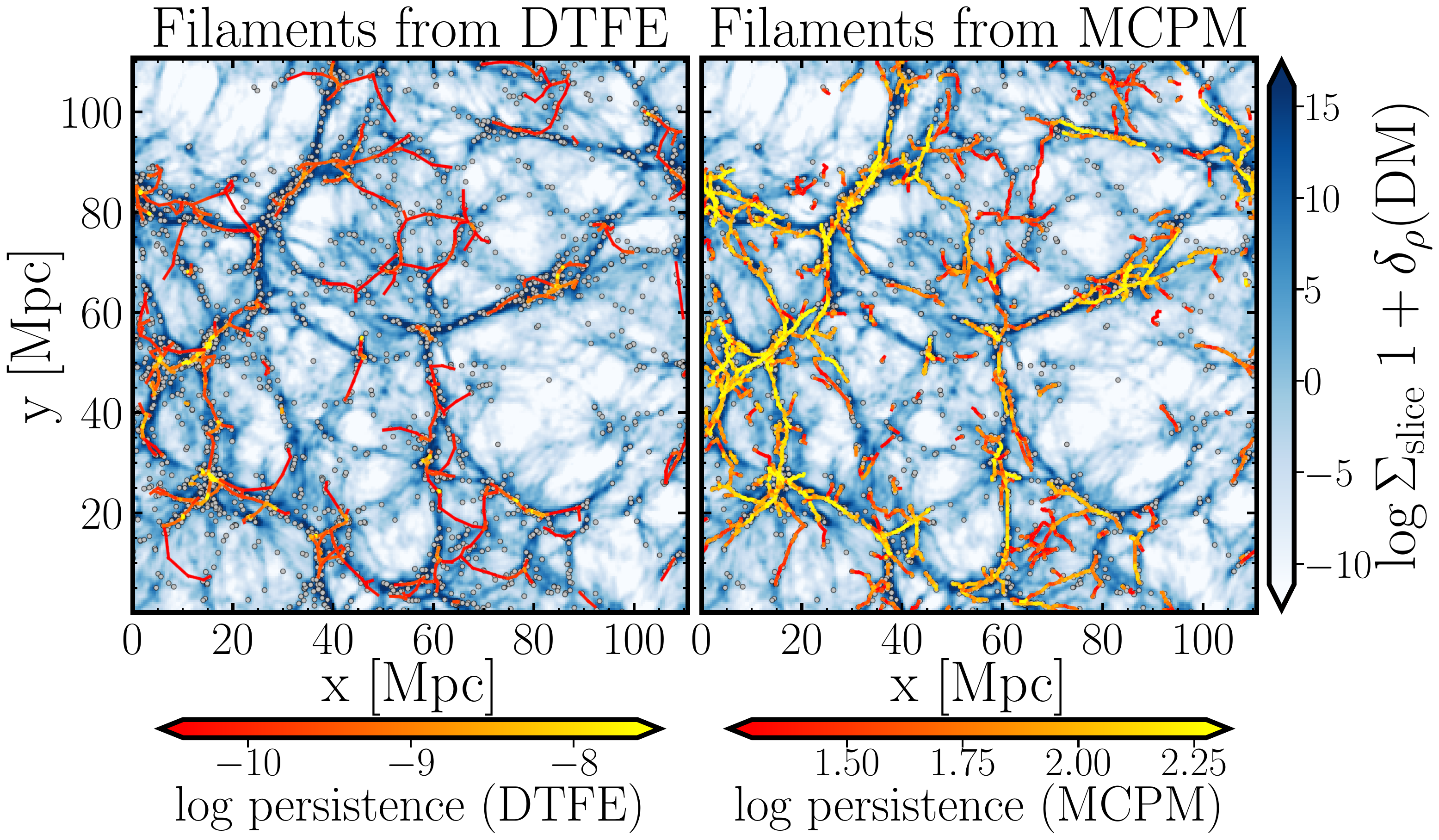}{0.98\textwidth}{}
}
\vspace{-25pt}
\caption{
A 2D visual comparison of the filaments identified by {\disperse} from the DTFE density field (left) and the MCPM density field (right) in the same 10 Mpc thick slices in TNG100 at $z=0$. In both panels, the underlying DM overdensity along the line of sight is represented by the blue-white colorbar, and gray circles represent galaxies. The individual filament segments from DTFE and MCPM are curves colored by the average persistence value of a segment (red is less persistent, yellow is more persistent). The MCPM density field identifies the filamentary structure with significantly higher fidelity, including the less prominent filaments that the DTFE density field method misses.
}
\label{fig:vis_fils}
\end{figure*}

After determining the best persistence and smoothing parameters for our two separate density field runs using {\disperse}, we compare the output filamentary structure in each. We begin by visually comparing the differences between filaments identified from the regular discrete DTFE density field and the new MCPM density field. In Figure~\ref{fig:vis_fils}, we present a 2D visualization of galaxies, the underlying DM density field, and the {\disperse}-identified filaments from the input of the DTFE density field (left) and the input of the MCPM density field (right). The same 10~Mpc thick slice is shown for the $z=0$ snapshot of TNG. In both panels, the individual filament segments are represented by curves colored by the average persistence value of the two endpoints of a segment (more robust filaments are colored yellow). Galaxies are represented by small gray circles and the DM overdensity, {\deltadm}, smoothed on a $256^3$ resolution scale and integrated along the 10~Mpc line of sight, is colored white-to-blue (with darker colors indicating higher overdensities).

The differences in the identified filamentary structure between the two density field inputs are quite dramatic. The MCPM density field run identifies many more filaments, faithfully tracing the underlying DM distribution. 
Recall that the persistence of an identified topological structure is a measure of how robust the structure is to local variations in the density field induced by shot noise. We can therefore liken the persistence of a filament segment to its ``detection confidence'' -- a more persistent filament segment is a more robust structure with respect to the shot noise in the input data.
In what follows, we use the terms ``persistence'' and ``prominence'' interchangeably. 
The DTFE density field run misses many of the less persistent filaments that its MCPM counterpart captures. Thus, DTFE only enables identifying the most persistent filaments that trace higher matter overdensities. We also see that the shapes of the MCPM-based filaments appear to be much more natural, i.e., more curved shapes with very few sharp line segments, than those of the DTFE-based filaments. Several DTFE-based filaments appear very sharp or unphysical (e.g., looped filaments or filaments that do not follow the spatial distribution of the DM matter density). 
We find similar differences in the reconstructions at higher redshifts.

\begin{figure*}[htbp]
\vspace{-10pt}
\centering
\gridline{\fig{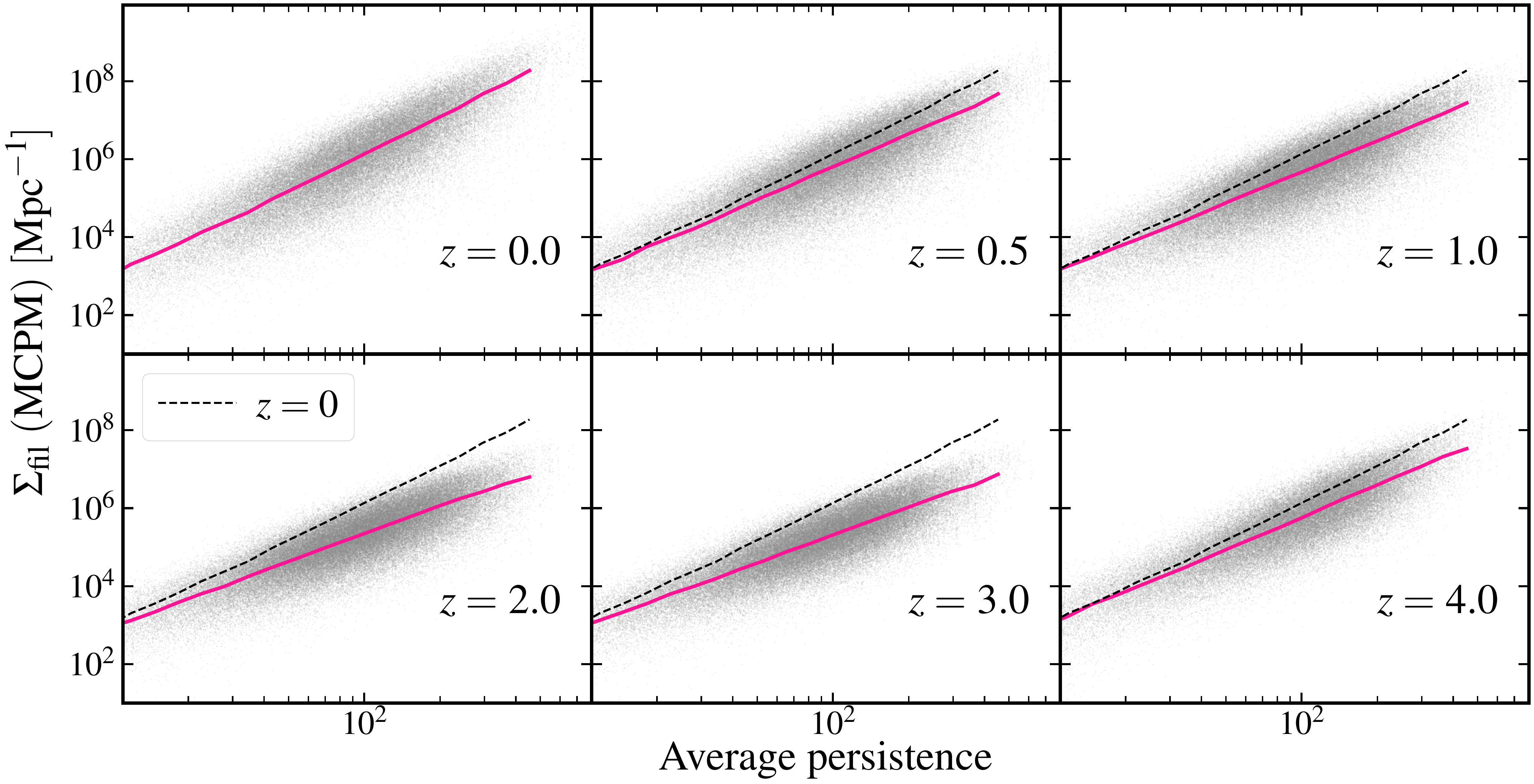}{0.8\textwidth}{}
}
\vspace{-25pt}
\caption{
The relationship between local filament line density (Eq.~\ref{eq:sigmafil}) and average persistence of a filament segment -- mean of the persistence of the two critical points at the two ends of a segment -- at different redshifts. The solid magenta line shows the running median relationship at that redshift and the dashed black line shows the $z=0$ relation.
We see a strong monotonic correlation between {\sigmafil} and average persistence of filament segments at all $z$.
}
\label{fig:sigmavpers}
\end{figure*}

\begin{figure}[htbp]
\centering
\gridline{\fig{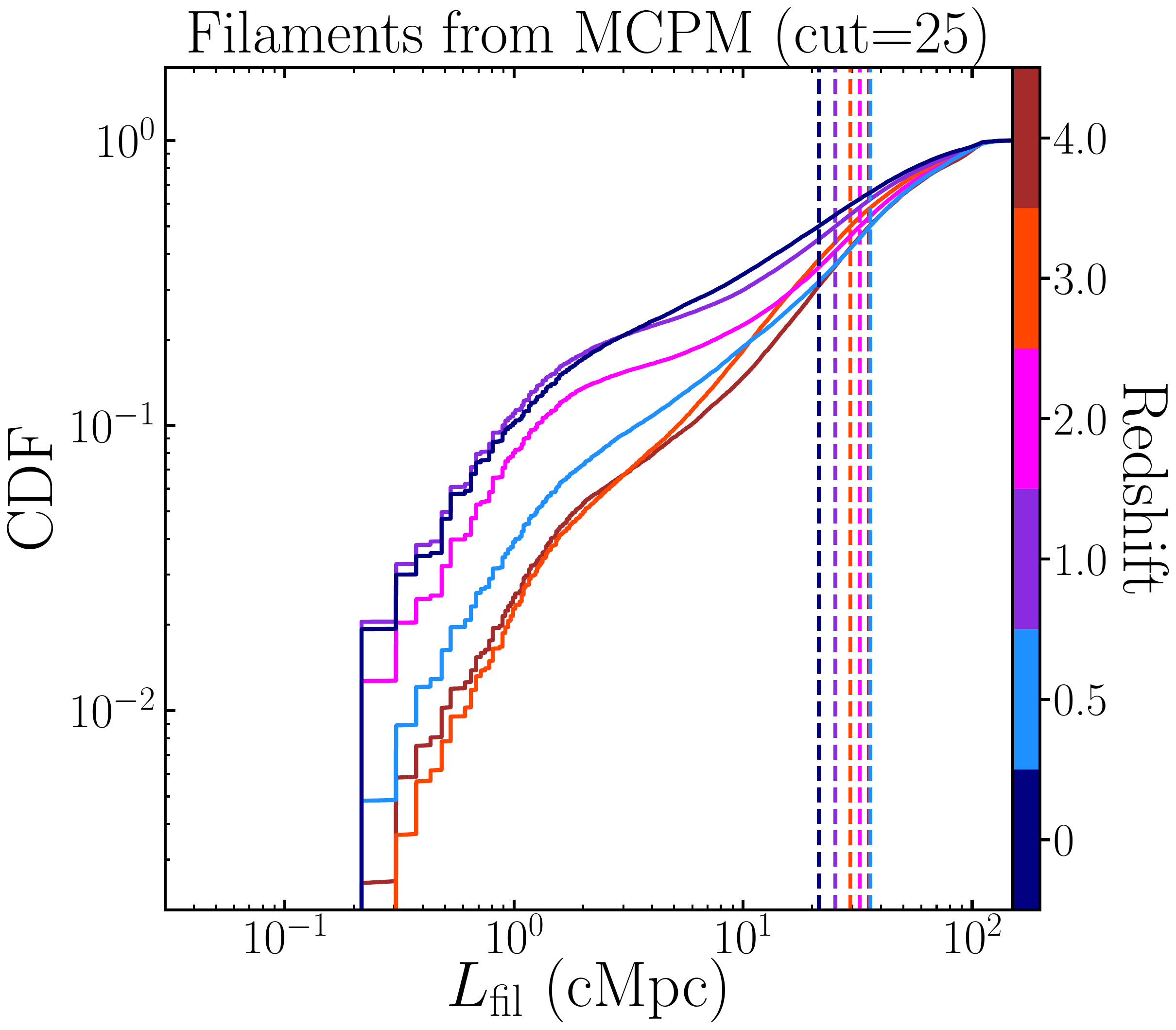}{0.35\textwidth}{}
}
\vspace{-25pt}
\gridline{\fig{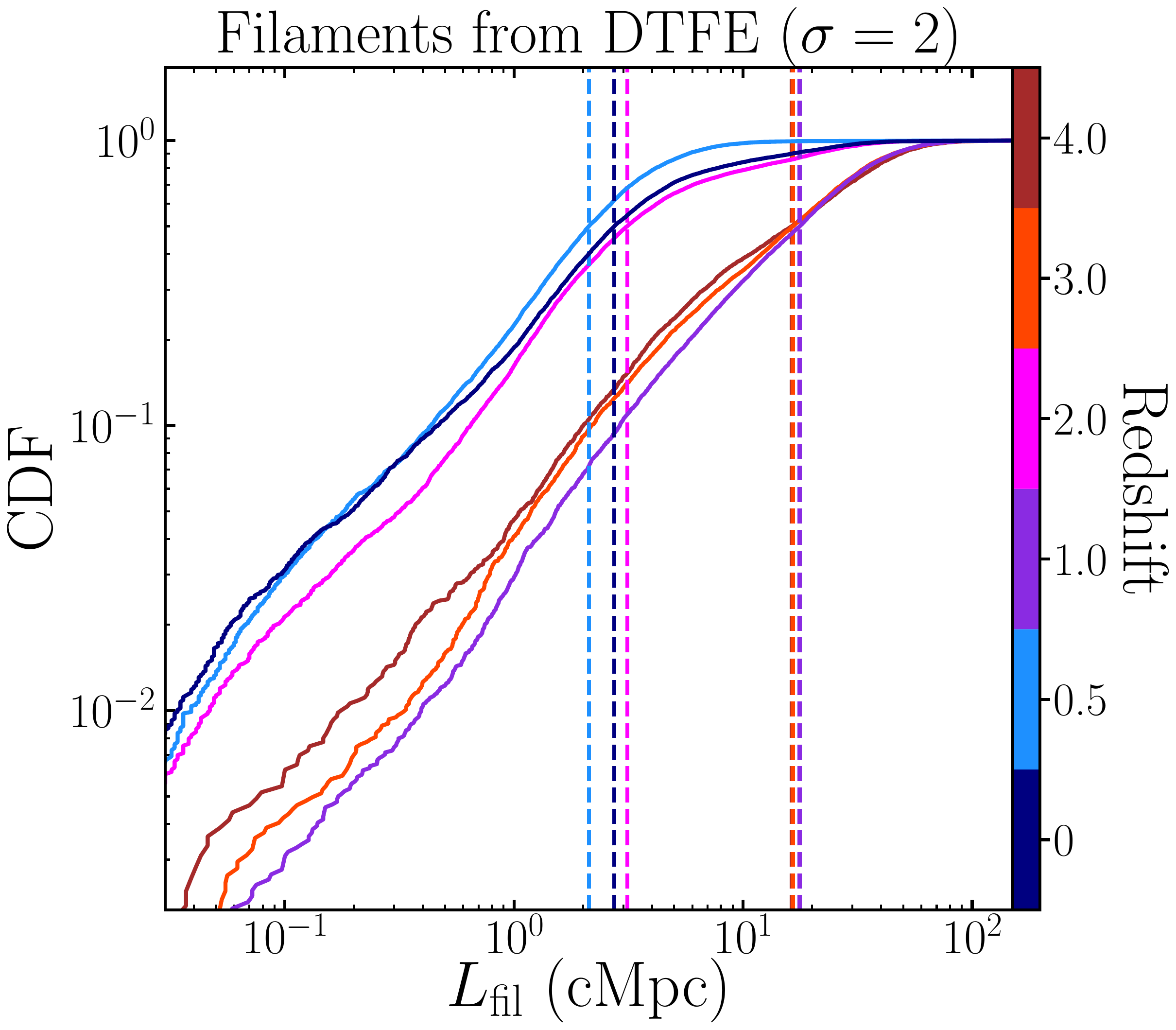}
{0.35\textwidth}{}
}
\vspace{-25pt}
\caption{
Cumulative distribution of total comoving length of filaments {\Lfil} from MCPM (top) and DTFE (bottom) at different redshifts (each with a different color). The vertical dashed lines represent the median {\Lfil} at a given $z$.
MCPM probes larger physical structures than DTFE at any given redshift.
}
\label{fig:lfil}
\end{figure}

Figure~\ref{fig:vis_fils} demonstrates
that both DTFE and MCPM runs identify filaments in the most overdense regions. These are the most prominent filaments around the most massive halos. However, MCPM also identifies orders of magnitude more of the {\it less prominent filaments} missed by DTFE, which nevertheless trace the cosmic matter distribution. These are the red colored curves on the right panel of Figure~\ref{fig:vis_fils}. 
The persistence of a filament segment is likely related to the local overdensity along the segment. 
Unfortunately, the {\disperse} formalism does not include a characterization of filament density, instead providing discrete point set outputs for locations of filament segments and critical points. 
We circumvent this problem by making use of the MCPM density. We define a new quantity, which we deem the local {\it filament line density}, or {\sigmafil},
\vspace{-5pt}
\begin{equation}
    {\sigmafil} = \frac{\sum_{seg}{\deltamcpm}}{L_{seg}} \, ,
\label{eq:sigmafil}
\end{equation}
where $\sum_{seg}{\deltamcpm}$ is the sum of the MCPM overdensity along a filament segment, and $L_{seg}$ is the physical length of the segment. 
{\sigmafil} is the summed overdensity per unit length of a filament segment and has dimensions of 1/length. We only consider voxels containing the spine of a filament in this measurement. 
This quantity is a descriptor of the {\it local} 1D line of the filament segments and can depart significantly from the global line density -- the density along {\it all} line segments making up a filament -- as filaments can be $\sim$100 Mpc long with significant variations in matter density along different segments. 

Past studies have used line density, filament width, or a similar metric to quantify filament thickness. \citet{Birnboim16}, in their analytical study of the hydrodynamic stability of filaments, analyzed the total filament mass per unit length. \citet{Zhu22} used Hessian-based estimates of the cosmic web structure to identify wide and narrow filaments (which they term ``prominent'' and ``tenuous'', respectively). 
Others, such as \citet{Ramsoy21} and \citet{GE22}, use simulation particle data to characterize typical density profiles and core widths of filaments. 
In contrast, the approach we adopt here is designed to apply to observational data, for which both the MCPM overdensity along a filament segment and the length of a segment can be computed \citep{Burchett20,Simha20,Wilde23}.

We plot the local filament line density as a function of the persistence of each filament segment at six different redshifts in Figure~\ref{fig:sigmavpers}. 
The magenta line shows the median of the relationship at that redshift, and the dashed black line shows the median relation $z=0$. {\sigmafil} correlates strongly with average persistence at each redshift (with Spearman correlation coefficients $>$0.8 in all cases). 
The filament line density increases monotonically with average persistence, indicating that the former is a good physical proxy for quantifying the prominence of filamentary structures. We note a large -- $\gtrsim$7 dex -- range in {\sigmafil}, indicating the vast dynamical range in cosmic densities that these structures probe. This figure confirms that less prominent filaments identified by MCPM do in fact have lower local density per length along the filament segment. Highly prominent filaments that both MCPM and DTFE identify have higher local line density.
With decreasing redshift, larger persistence values are seen, caused by high-persistence structures generated in increasingly high-density environments as groups and clusters take shape later in the universe.
At higher $z$, the median {\sigmafil}-persistence relationship is below the $z=0$ relation and becomes somewhat shallower, but the qualitative trend remains unchanged.

At this juncture, we crucially remind the reader that the above visualizations show  only a {\it finite slice} of the whole volume, projected on to a 2D plane. In other words, projection effects will make some of these filaments appear very small because we are slicing at $\lesssim$10\% of the volume in the z-direction and show filament segments that fall entirely within the slice.
In fact, virtually all of the MCPM filaments, including those with low-prominence, are large structures of length on the order of Mpc or larger.

\begin{figure*}
\vspace{-10pt}
\centering
\gridline{\fig{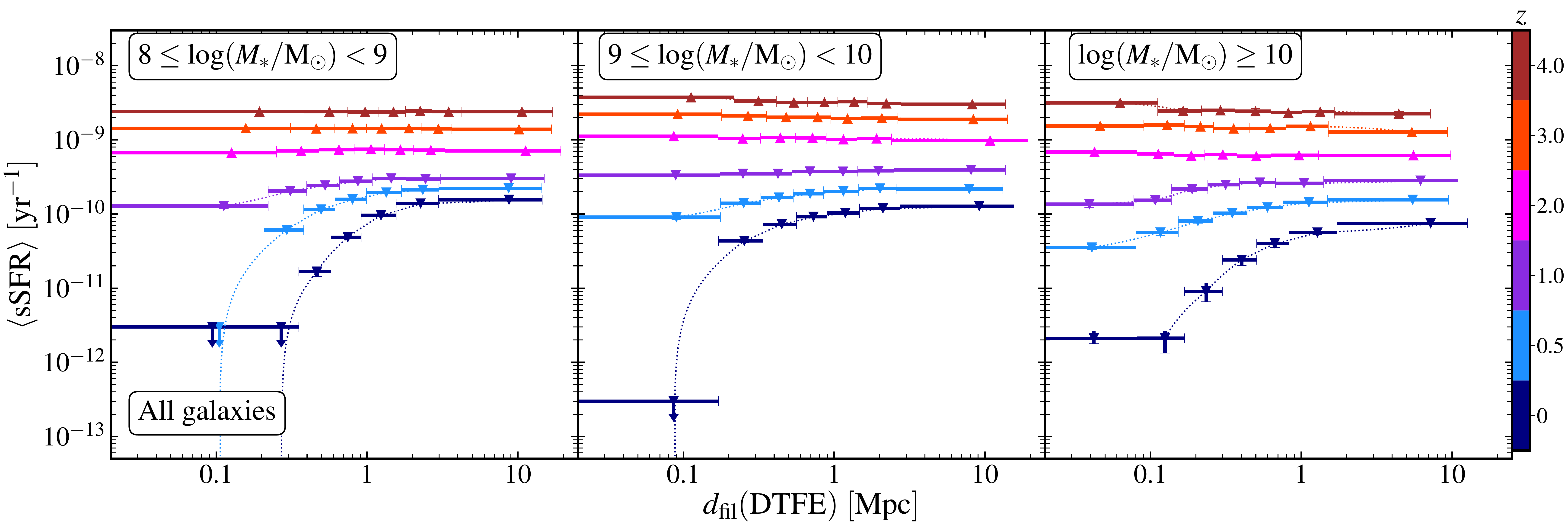}{0.88\textwidth}{}
} \vspace{-28pt}
\gridline{\fig{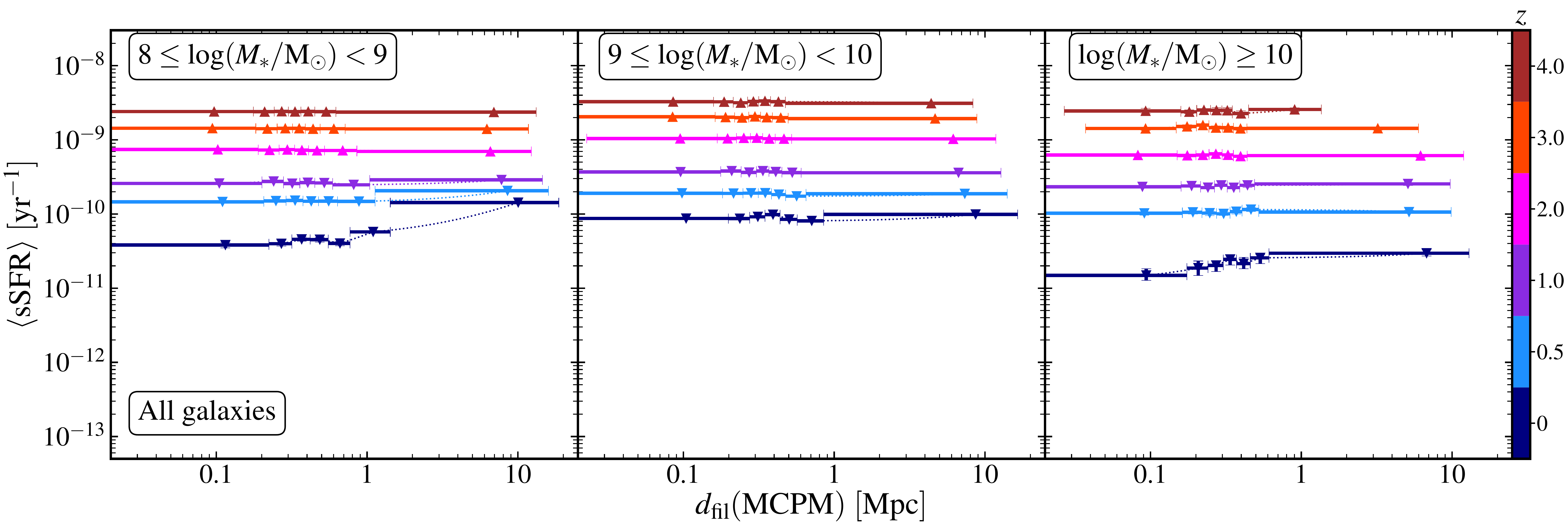}{0.88\textwidth}{}
} \vspace{-25pt}
\caption{
The median sSFR, {\medssfr}, as a function of distance to the closest filament identified from DTFE (top row) and MCPM (bottom row) for all galaxies of mass {\lowmsrange} (left), {\midmsrange} (middle), and {\highmsrange} (right) at redshifts $z=0$, 0.5, 1, 2, 3, and 4 as indicated by the colorbar on the right.
We find very little correlation between {\medssfr} and {\dfil}(MCPM), relative to {\dfil} (DTFE), in almost all cases.
}
\label{fig:medssfr_dfil}
\end{figure*}

To show the scales of the structures being probed in both the MCPM and DTFE filament catalogs, we present CDFs of the comoving filament lengths {\Lfil} from MPCM (top) and DTFE (bottom) at redshifts $z=0$, 0.5, 1, 2, 3, and 4 in Figure ~\ref{fig:lfil}. 
Vertical dashed lines indicate the median {\Lfil} values at the given redshifts. MCPM-based filaments are longer than DTFE-based filaments at all times and have a much smaller dynamic range in the medians -- between $\sim$20 and $\sim$45 Mpc from $z=0$ to 4. 
In contrast, the median DTFE-based filament ranges between $\sim$2 and $\sim$18 Mpc in length, ranging over almost an order of magnitude. 
Furthermore, $\sim$2-12\% of MCPM filaments have ${\Lfil}<1$ Mpc, but the fraction rises up to $\sim$25\% for DTFE filaments. 
MCPM filaments are therefore larger in physical size and probing larger-scale structure than DTFE filaments.

Our visual and statistical comparison of the MCPM and DTFE filament catalogs has led us to conclude that the accuracy and fidelity of identifying the structure of the cosmic web is improved when we replace the DTFE density field with the MCPM density field in the {\disperse} framework. 
There are several reasons to support our conclusion. 
(1) The MCPM overdensity {\deltamcpm} -- smoothed at any resolution -- correlates more strongly and monotonically with the underlying cosmic matter density field as traced by {\deltadm}, than the DTFE overdensity {\deltadtfe}.
(2) Filaments identified from the MCPM density fields trace out the DM density field 
much more accurately and completely than those identified from the DTFE density fields. This includes MCPM finding less persistent filaments that should nevertheless exist in lower-density -- but still overdense -- regions of the cosmic web. {\it An entire population of low persistence or low line density filaments is missed by the DTFE-based reconstructions}.
(3) MCPM filaments are much larger in physical size than DTFE filaments and the former contain a much smaller fraction of possibly spurious filaments with lengths close to the resolution scale of the density fields of the reconstruction. 
(4) Many DTFE filaments appear to have sharp/unphysical shapes, while MCPM filaments do not suffer from this issue.
We discuss the possible reasons for these improvements in Section~\ref{sec:improve}.


\section{The Effect of Filaments on Galaxy Evolution}
\label{sec:results}


\subsection{How Filaments Affect Star Formation}

We now study how the MCPM density field-based filaments of the cosmic web environment affect galactic star formation activity and gas fraction. The relationship is analyzed by selecting galaxies based on their stellar mass, redshift, and central/satellite status of galaxies.

\vspace{-5pt}

\subsubsection{sSFR vs. Distance to Filaments}
\label{sec:ssfrvdfil}

\begin{figure*}[htbp]
\vspace{-10pt}
\centering
\gridline{\fig{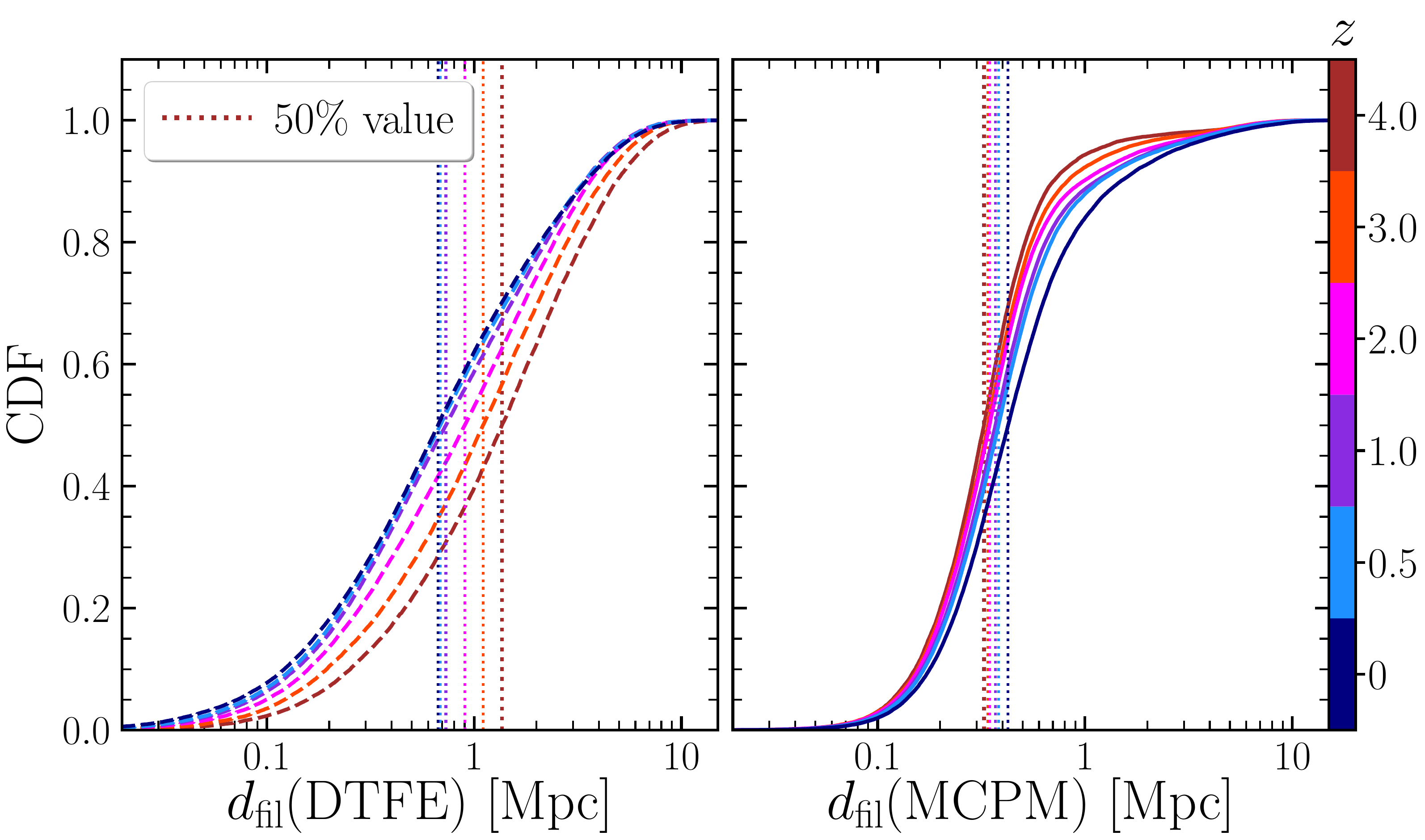}{0.7\textwidth}{}
}
\vspace{-25pt}
\caption{
Cumulative distributions of {\dfil}(DTFE) and {\dfil}(MCPM) at different redshifts. For MCPM filaments, most galaxies live very close to a filament spine at any redshift, while many galaxies live relatively far from DTFE filaments. 
}
\label{fig:dfilcdf}
\end{figure*}

First, we study the effect of filaments on the specific star formation rate (sSFR) of galaxies. To quantify the filamentary environment, we calculated the comoving transverse distance from each galaxy to the spine of the nearest identified filament, {\dfil}, similar to \citetalias{Hasan23}. We measured this distance for both the DTFE and MCPM methods, denoted as {\dfil}(DTFE) and {\dfil}(MCPM), respectively. 
For each of redshifts $z=0$, 0.5, 1, 2, 3, and 4, we separate galaxies into three bins of stellar mass: {\lowmsrange}, {\midmsrange}, and {\highmsrange} and seven bins of {\dfil}(DTFE) and {\dfil}(MCPM). We then calculate the median sSFR, {\medssfr}, for each bin of {\dfil}(DTFE) and {\dfil}(MCPM) for a given mass range and redshift.
The binning of {\dfil} is such that there are approximately equal numbers of galaxies in each bin in order to maximize the statistical significance of each bin.

The results are presented in Figure~\ref{fig:medssfr_dfil}. The top panels show {\medssfr} as a function of {\dfil}(DTFE), and the bottom panels show {\medssfr} as a function of {\dfil}(MCPM) for all galaxies. The left, middle, and right panels represent galaxies of stellar mass {\lowmsrange}, {\midmsrange}, and {\highmsrange}, , respectively. The different sets of colored data points,  error bars, and dotted curves represent {\medssfr}, $\pm1\sigma$ bootstrapped errors on {\medssfr}, and spline interpolations of the {\medssfr}-{\dfil} relation, respectively, at different redshifts. 
As in \citetalias{Hasan23}, some data points are presented as upper limits due to the minimum resolvable SFR$=10^{-2.5}$~{\Msunyr} in TNG100 \citep{Terrazas20}. 

There is a striking difference between the MCPM and DTFE methods in terms of the dependence of the average star formation on {\dfil}. When DTFE-based filaments are considered, galaxies quench as they get closer to filaments at low redshift for all stellar masses. This is especially true at $z=0$, where a decrease in $\gtrsim$ 2 dex in sSFR occurs from high to low {\dfil} at any mass. In contrast, for MCPM-based filaments, there is virtually no dependence of {\medssfr} on {\dfil}, save for the {\lowmsrange} and {\highmsrange} galaxies at $z=0$. Otherwise, when MCPM-based filaments are considered, the distance to filaments appears to have almost {\it no effect} on star formation of high, intermediate, or low mass galaxies. 
This MCPM result also differs from that of \citetalias{Hasan23}, where we used the DTFE density field to identify filaments. For both DTFE and MCPM filaments, there is no statistical
correlation between median star formation and distance to filaments at $z\geq2$.

The difference in our measured statistics between the two methods can be explained by the difference in the identified filamentary structure. We saw in Section \ref{sec:dtfevmcpm} that most galaxies reside close to MCPM filaments, whereas many galaxies are not close to DTFE filaments. We show this quantitatively in Figure~\ref{fig:dfilcdf}, which compares the CDFs for {\dfil}(DTFE) and {\dfil}(MCPM) at different redshifts. The vertical dashed line represents the median {\dfil} at a given redshift. On average, galaxies are significantly closer to the spine of a filament identified from MCPM than one identified from DTFE. The median {\dfil}(MCPM) is approximately 0.4 Mpc in all redshifts, while the median {\dfil}(DTFE) is $\sim$0.7--1.5 Mpc. 90\% of galaxies are within 1--1.5 Mpc (depending on redshift) of an MCPM filament, indicating that {\it virtually all galaxies live close to a filament spine}.  
When filaments are identified with DTFE, a substantial fraction ($\approx$20--40\%) of galaxies are found ${\dfil}\gtrsim2$~Mpc from filament spines.

We note that by analyzing the relationships to nearby filament spines, we so not explicitly classify galaxies as belonging to particular structures, as have other studies in the literature. 
Some of the existing reconstruction methods segment space into filaments, nodes, sheets, and voids based on some smoothing scale, while {\disperse} gives discrete estimates of critical points that are linked together to form 1D filament line segments. 
Studies such as those of \citet{Cautun14} or \citet{GV19} explicitly classify halos/galaxies as belonging to filament, node, sheet, or void environments. In our analyses, we do not attempt any such classification schemes. Instead, we describe the cosmic web environment of a galaxy in terms of the nearest filament spine. This allows for a fairly continuous representation of the cosmic web environment, as opposed to discrete definitions of galaxies either belonging to a filament or not. Therefore, many of the galaxies that reside close to filament spines are likely associated with nodes or massive halos (and effect we account for in section~\ref{sec:massive}) or even sheets which appear to be filaments in 2D projection \citep[e.g.,][]{disperse1,Cautun14}. 
As outlined in \citet{Libeskind18}, the different filament identification methods were designed with different scientific goals in mind, and even when applied to the same dataset, they sometimes show significant discrepancies in the statistics of output cosmic web structure.

\begin{figure*}
\vspace{-10pt}
\centering
\gridline{\fig{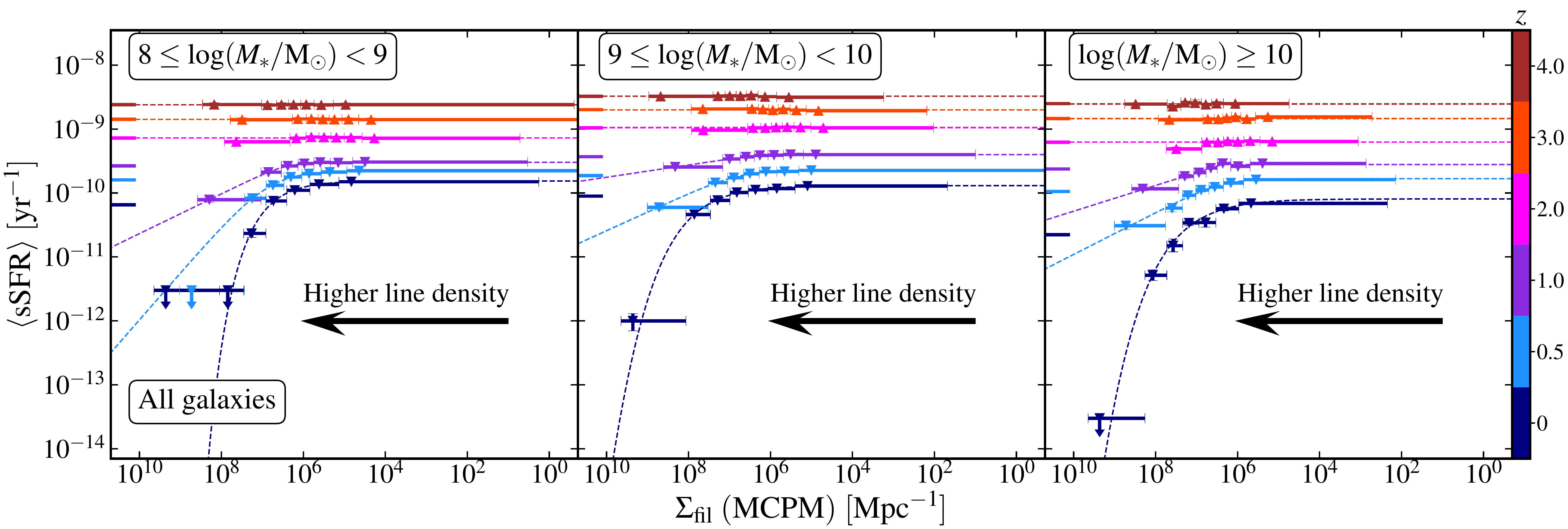}{0.93\textwidth}{}
} \vspace{-25pt}
\gridline{\fig{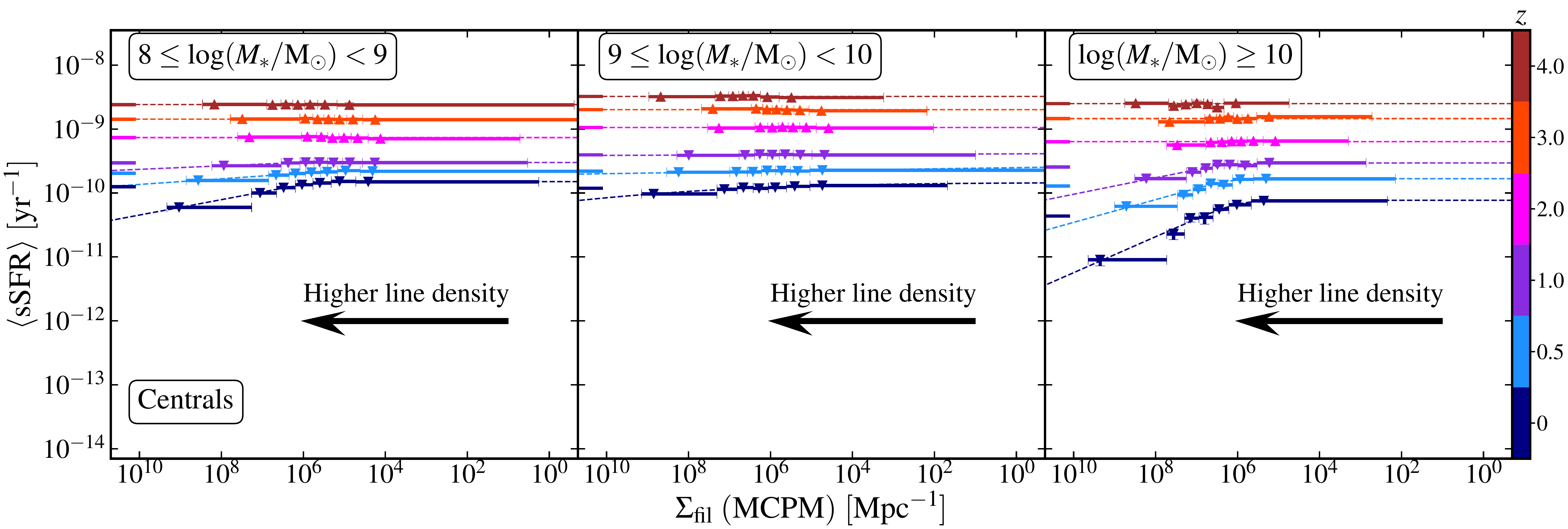}{0.93\textwidth}{}
} \vspace{-25pt}
\gridline{\fig{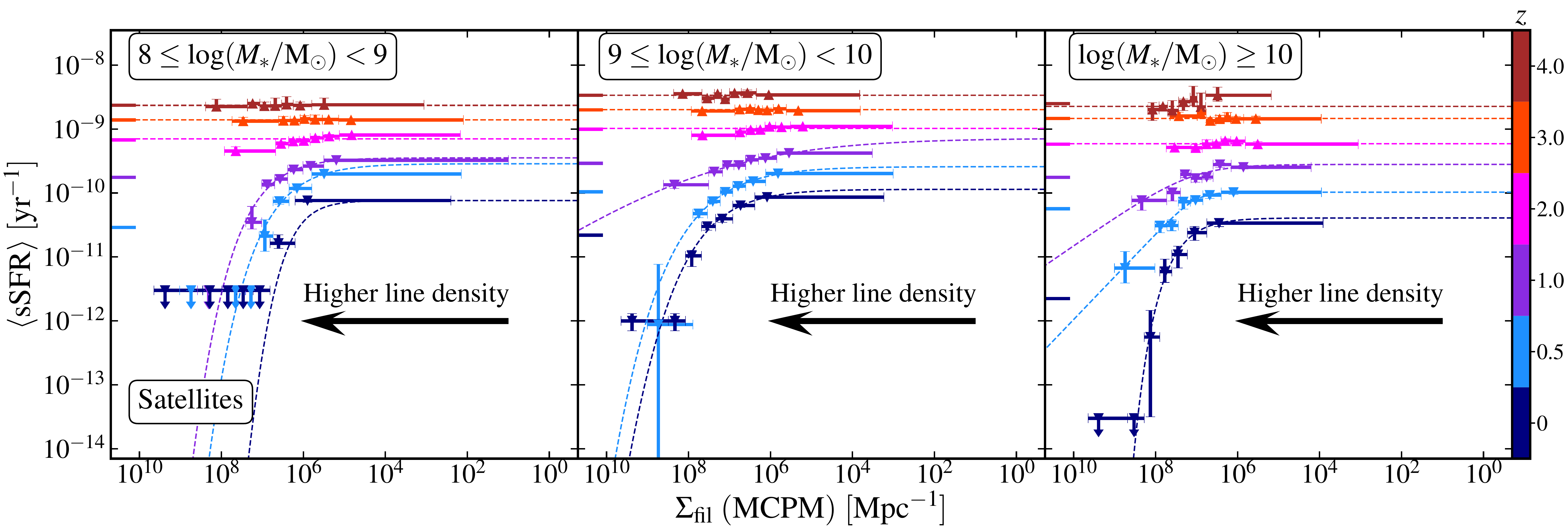}{0.93\textwidth}{}
} \vspace{-20pt}
\caption{
{\medssfr} as a function of the local filament line density {\sigmafil} (Eq.~\ref{eq:sigmafil}) for all galaxies (top row), central galaxies (middle row) and satellite galaxies (bottom row). The mass ranges and redshifts are the same as in Figure~\ref{fig:medssfr_dfil}. The overall median sSFR for a given {\ms} and $z$ is indicated by short horizontal solid lines on the left edge of each panel. Dashed curves represent the best-fit {\medssfr}-{\sigmafil} relationships to Eq.~\ref{eq:betamodel}. {\it At lower redshifts, galaxies near high line density filaments are quenched, while galactic star formation is not suppressed near low line density; at higher redshifts, line density does not affect galactic star formation}.
}
\label{fig:medssfr_sigmafil}
\end{figure*}

\vspace{-5pt}

\subsubsection{sSFR vs. Filament Line Density}
\label{sec:ssfrvsigmafil}

The transverse distance to a filament is not an adequate descriptor of cosmic web environment when MCPM filaments are considered. Our findings motivate the need for a new metric by which to quantify the cosmic web environment that takes into account the effect of the local line density along the filament segment closest to a galaxy. 
Therefore, we repeat our calculations of {\medssfr} at different masses and redshifts as functions of local filament line density {\sigmafil}, as defined in Eq.~\ref{eq:sigmafil}, of the nearest filament segment.

In Figure~\ref{fig:medssfr_sigmafil}, we present {\medssfr} for low, intermediate, and high {\ms} galaxies (different columns) and in six bins of redshift (different colored points) as functions of {\sigmafil} of the nearest filament segment for all galaxies (top panels), central galaxies only (middle panels) and satellite galaxies only (bottom panels). 
We select centrals as the most massive galaxies in their host DM halos, and all other galaxies are satellites.
As in Figure~\ref{fig:medssfr_dfil}, data points and error bars represent {\medssfr} and their bootstrapped errors $\pm1\sigma$, and upper limits are as described above. To be consistent with Figure~\ref{fig:medssfr_dfil}, the line density increases to the left-hand side of each panel (indicated by an arrow in each panel).
The discrete colorbar on the right of each row indicates the redshift. 
On the left edge of each panel, horizontal solid lines indicate the {\medssfr} for all galaxies at that mass and redshift, regardless of filamentary environment, as a reference to gauge the degree of quenching relative to that mass and redshift.
Similar to Figure~\ref{fig:medssfr_dfil}, the binning of {\sigmafil} is such that there are approximately equal numbers of galaxies in each bin (typically on the order of 100). We verify that changing the number of bins does not affect our results.

We find that galaxies are generally quenched when they live near high-line density filaments at later times. Low-mass galaxies are the most strongly affected: even at $z=1$, there is almost an order of magnitude suppression of star formation in galaxies near high line density filaments (${\sigmafil}\gtrsim10^8~\mathrm{Mpc}^{-1}$). At $z=0$, galaxies of all masses are quenched if they live near such high line density filaments. 
This effect of reduced star formation activity in galaxies near high line density filaments decreases with increasing redshift, but still persists out to $z\simeq1$ for all masses.
At $z\geq2$, there is virtually no dependence of {\medssfr} on {\sigmafil} for any galaxies, even for the same value of {\sigmafil}. For example, from ${\sigmafil}\sim10^6~\mathrm{Mpc}^{-1}$ to ${\sigmafil}\sim10^9~\mathrm{Mpc}^{-1}$, {\medssfr} drops by an order of magnitude or more for galaxies of all masses at $z\leq0.5$, but for such line densities, there is no change in median star formation of galaxies at $z\geq2$.

For lower-mass ({\mslow}) galaxies, satellites almost exclusively drive the {\sigmafil}-dependence of star formation activity, as shown in the bottom panels of Figure~\ref{fig:medssfr_sigmafil}. 
The dependence of star formation of {\mslow} satellites on filament line density is seen out to $z\sim2$. 
{\mslow} satellites in high line density filaments stop forming stars at $z\leq1$, while centrals of the same mass experience a much smaller reduction in {\medssfr} at $z=0$.  
For {\highmsrange} galaxies, on the other hand, both centrals and satellites feel the effect of local filament line density. {\highmsrange} satellites near high line density filaments experience a $>$1 dex drop in {\medssfr} at $z\leq0.5$, and high-mass centrals see such a drop in {\medssfr} at $z=0$ (with the effect at $z=0.5$ being less strong but still quite significant). 
Regardless of the low-redshift trends, the {\sigmafil}-dependence of star formation activity for the massive centrals or satellites disappears at $z>2$.
We note that the lower number of {\highmsrange} satellites at $z\geq2$ is a significant source of uncertainty.

The dashed curves in each of the plots above represent the best-fit relationship between {\medssfr} and {\sigmafil} for the following functional form,
\vspace{-5pt}
\begin{equation}
    Y(X) = Y_0 \bigg(1+\bigg(\frac{X}{X_0}\bigg)^{\alpha} \bigg)^{\beta} \, ,
\label{eq:betamodel}
\end{equation}
where $Y={\medssfr}$, $X={\sigmafil}$, 
$X_0$ is the {\sigmafil} value where the {\medssfr} starts to decline from its characteristic value $Y_0$, and $\alpha$ and $\beta$ are power-law exponents. This is equivalent to the generic $\beta$ model that has historically been used to model the gas density profile of galaxy clusters \citep[e.g.,][]{CF78}, but has also been applied in the context of gas density and galaxy density profiles in filaments \citep[e.g.,][]{Bonjean20,GE22}. In Eq.~\ref{eq:betamodel}, we substitute the radius typically used in these models for the filament line density.
While we have adopted this functional form, we note that some other models, such as the generalized NFW \citep[e.g.,][]{Nagai07} and Einasto models \citep[e.g.,][]{Einasto65}, may also describe the relationships that we see quite well.
In any case, the functional form we choose describes our qualitative trends well: {\medssfr} stays roughly constant until a truncation point in {\sigmafil}, above which it drops as a power law described by some exponent. 
We find that $\alpha\simeq1$ in most cases, for which Eq.~\ref{eq:sigmafil} can be approximated as a three-parameter model. 
At $z\geq2$, $\beta$ is approximately zero in all cases, for which {\medssfr} is virtually invariant with {\sigmafil}.

Although we showed that virtually all galaxies live close to MCPM filaments, we also applied {\dfil}(MCPM) cuts on the {\medssfr}-{\sigmafil} relationships above. Choosing only galaxies within 1 or 2 Mpc of MCPM filaments, we recomputed {\medssfr} as a function of {\sigmafil}. As expected, we do not observe qualitative changes in the {\medssfr}-{\sigmafil} relationship for any $z$ or {\ms}.

\subsubsection{Quenched Fraction vs. Filaments}

\begin{figure*}[htbp]
\vspace{-10pt}
\centering
\gridline{\fig{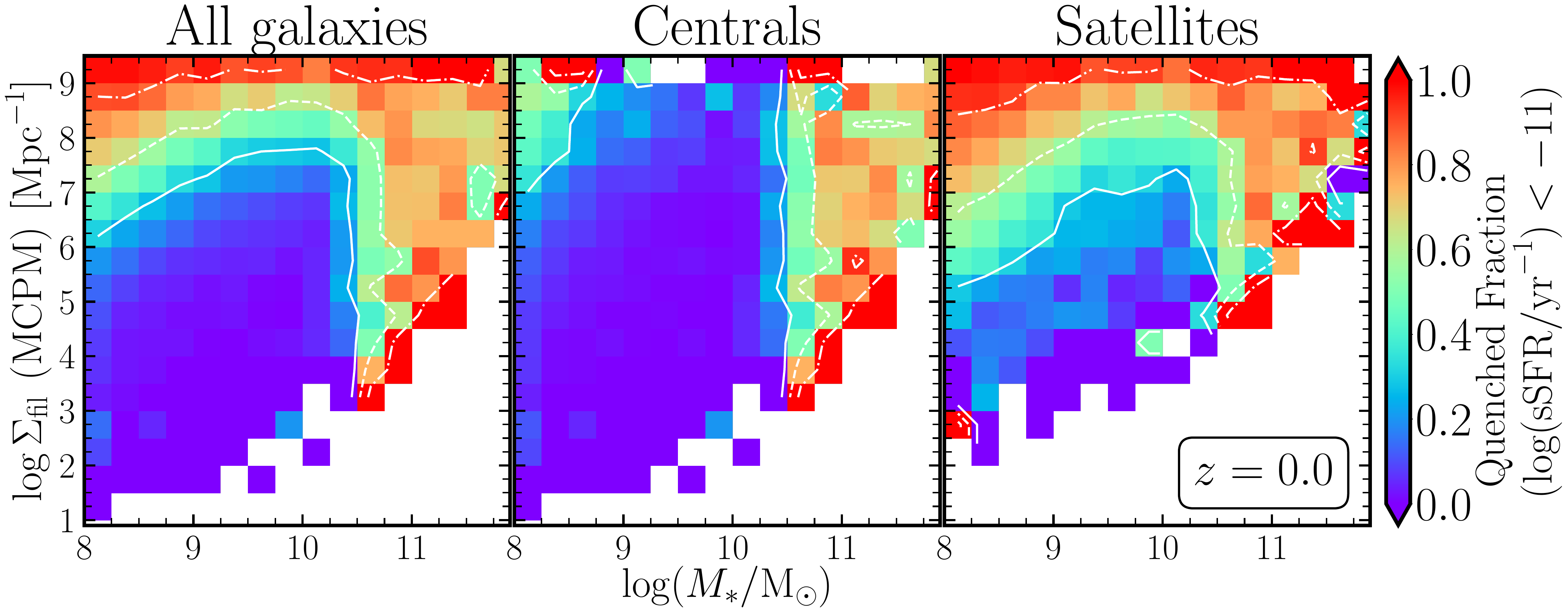}{0.9\textwidth}{}
}
\vspace{-25pt}
\gridline{\fig{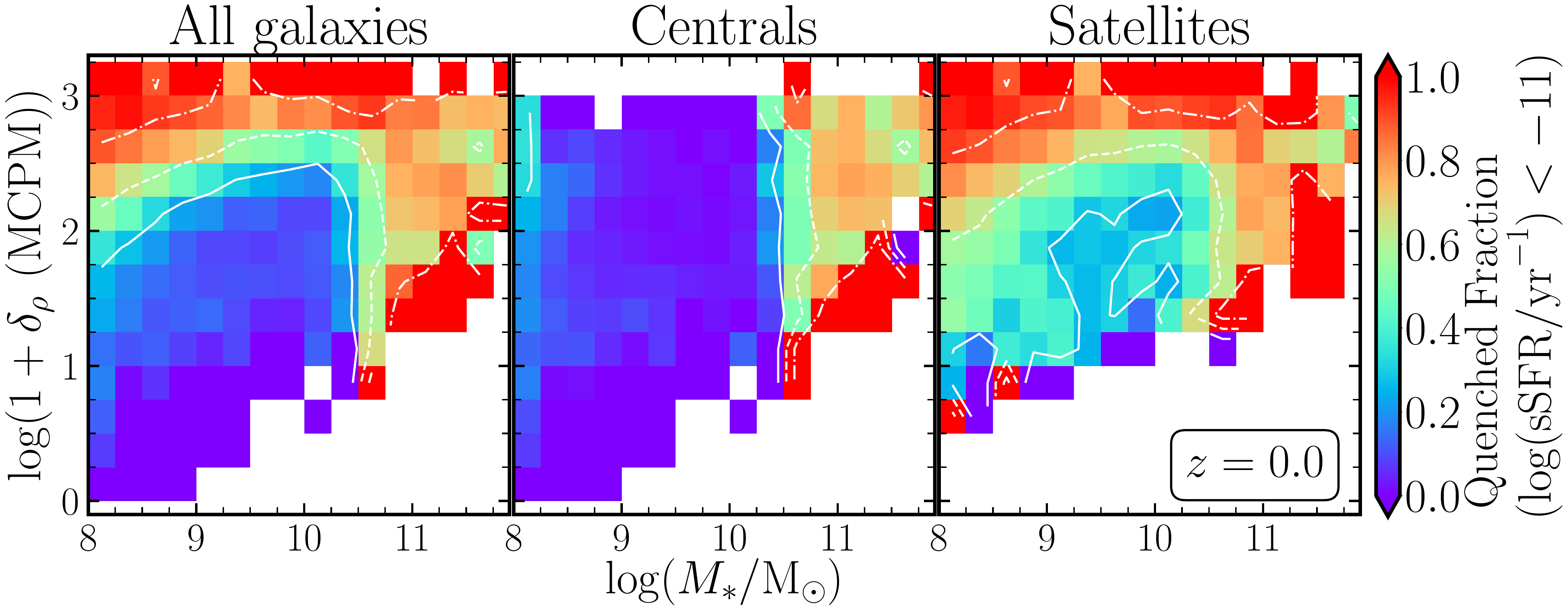}{0.9\textwidth}{}
}
\vspace{-25pt}
\caption{
The median quenched fraction, i.e., the fraction of galaxies with $\log(\mathrm{sSFR}/\mathrm{yr}^{-1})<-11$, at $z=0$ as a function of stellar mass and filament line density (top row) and MCPM overdensity (bottom row), for all galaxies (left), centrals (middle), and satellites (right). Solid, dashed, and dot-dashed curves represent contour lines enclosing the \nth{30}, \nth{60}, and \nth{90} percentiles of the distribution, respectively. At high masses, most galaxies are quenched regardless of the filament line density or local overdensity, low-mass galaxies quench at lower line density than higher-mass galaxies. Satellites are strongly affected by the environment, whereas centrals are hardly affected. 
}
\label{fig:qf}
\end{figure*}

Median sSFR is one indicator of the star formation activity of a population, but many others are used in the literature. Here, we assess the cosmic web dependence of the quenched fraction, the fraction of galaxies with sSFR below a certain threshold, as well as 
the red fraction, the fraction of galaxies with photometric colors above a certain threshold.
Similar to the observational studies of \citet{Peng10} and \citet{Peng12}, we study the joint dependence of quenched and red fractions of galaxies on mass and environment to understand the relative importance of internal vs. external properties of galaxies in quenching them. Unlike these studies, however, we quantify the environment using local filament line density as we defined above.
We define the quenched fraction as the fraction of galaxies with $\log(\mathrm{sSFR}/\mathrm{yr}^{-1})<-11$, based on the commonly used criterion to separate star-forming and quenched galaxies in the local universe \citep[e.g.,][]{Bluck20,Donnari21b}. 
We define the red fraction as the fraction of galaxies with $g-r>0.6$ (where $g$ and $r$ are SDSS photometric filters; \citet{Stoughton02}). We adopted this color cut based on (1) the separation of red sequence and blue cloud galaxies in the well-known galaxy color bimodality observed at low redshift \citep[e.g.,][]{Kauffmann04}, and (2) the excellent agreement between observed and simulated $g-r$ colors at $z<0.1$, as well as the separation of red and blue galaxies for TNG100 at $g-r>0.6$ reported by \citet{Nelson18}.

We bin galaxies in stellar mass and filament line density and measure the median quenched fraction as defined above in each 2D {\ms}-{\sigmafil} bin. 
The median quenched fraction at $z=0$ is presented for all galaxies (left), centrals (middle), and satellites (right), as a function of {\ms} and {\sigmafil} in the top row of Figure~\ref{fig:qf}. In each panel, solid, dashed, and dashed-dotted contours enclose 30\%, 60\%, and 90\% of the distribution, respectively. 
For completeness, we also repeat this analysis in terms of MCPM overdensity {\deltamcpm}, itself a measure of the local environment, and present the quenched fraction as a function of {\ms} and {\deltamcpm} in the bottom row of Figure~\ref{fig:qf}.

It is apparent that the environment is not an important factor in quenching galaxies at ${\logms}>10.5$. Most galaxies are quenched above this mass, regardless of their nearby filament, or local overdensity.
This is primarily a consequence of the AGN feedback model in TNG, which is known to quench galaxies very rapidly as their central BHs reach a mass of $\sim10^{8.2}$~{\Msun} \citep[e.g.,][]{Terrazas20,Zinger20}. 
In the highest-line density filaments (${\sigmafil}\gtrsim10^8~\mathrm{Mpc}^{-1}$), the vast majority of galaxies are quenched, albeit lower-mass centrals are star-forming even in these environments.  
At ${\logms}<10.5$, for a fixed mass, the quenched fraction increases with increasing {\sigmafil} or {\deltamcpm}, implying that lower mass galaxies can quench near lower line density filaments than higher mass galaxies. This is driven by satellites, which transition from star-forming to quenched at very low {\sigmafil}. 
For a fixed {\sigmafil} or {\deltamcpm}, low- and intermediate-mass satellites are much less star-forming on average than low- and intermediate-mass centrals. 
We find that most centrals at ${\logms}<10.5$ are star-forming, regardless of {\sigmafil}.
We verify that changing the definition of the quenching threshold, e.g., to 0.5 or 1 dex below the star-forming main sequence, does not qualitatively impact our results.

We repeat our analyses in terms of the red fraction as defined above and found that the results are almost identical, showing that galaxies being quenched or red are virtually equivalent phenomena.
That most ${\logms}>10.5$ galaxies are red regardless of their environment is in agreement with \citet{Peng10}, despite slightly different definitions of the red fraction and environmental density measures (they used a galaxy number density).
However, \citet{Peng10} reported that, for lower-mass galaxies, the ridge-line separating mostly star-forming and mostly quenched galaxies is approximately flat, or even mildly decreasing in local density, with increasing {\ms}. In contrast, we find that this ridge-line between star-forming and quenched low-mass galaxies moves to higher filament line density and local overdensity with rising {\ms}. 
This may offer some important physical insights about the relative effect of environment on quenching.

Our result that satellites are the most susceptible to the environment is in line with the findings of \citet{Peng12}. 
Furthermore, our results are broadly consistent with the findings of \citet{Geha12} from SDSS DR8 that in the local universe, almost all the quenched galaxies with ${\logms}<9$ are satellites or centrals near massive neighbors and that virtually all centrals of this mass are star-forming in low-density environments.
Our qualitative results here do not change if we adopt slightly different cuts for the red fraction, e.g., $g-r>0.55$ or $g-r>0.65$.

Qualitatively, there is little change in the quenched or red fraction as a function of filament line density or MCPM overdensity, except in one or two cases. For a fixed mass at low {\deltamcpm}, there is a substantial fraction of star-forming satellites, but for a fixed mass at low {\deltamcpm} there is a much smaller fraction of star-forming satellites.
This implies a non-monotonic relation between {\deltamcpm} and {\sigmafil} of filaments closest to low-mass satellites in low-density environments.

Aside from constraining the relative effect of mass and environment in quenching galaxies in the local universe, the results in this section also demonstrate how the filament line density {\sigmafil} is an effective metric to quantify the environmental dependence of quenching. It is (1) a more precise and detailed description of the surrounding galaxy environment, (2) a viable alternative to local environmental measures characterized by the distance to $N^{\mathrm{th}}$ nearest neighbor or galaxy number density ubiquitously used in the literature, and (3) a potentially powerful new way to constrain external physical quenching mechanisms.
Our results suggest that massive galaxies are quenched regardless of environment, but that a higher-density environment enables quenching in low-mass galaxies for a fixed stellar mass.

\vspace{-5pt}

\subsection{How Filaments Affect Gas Fraction}
\label{sec:fgasvfil}

\begin{figure*}
\centering
\vspace{-10pt}
\gridline{\fig{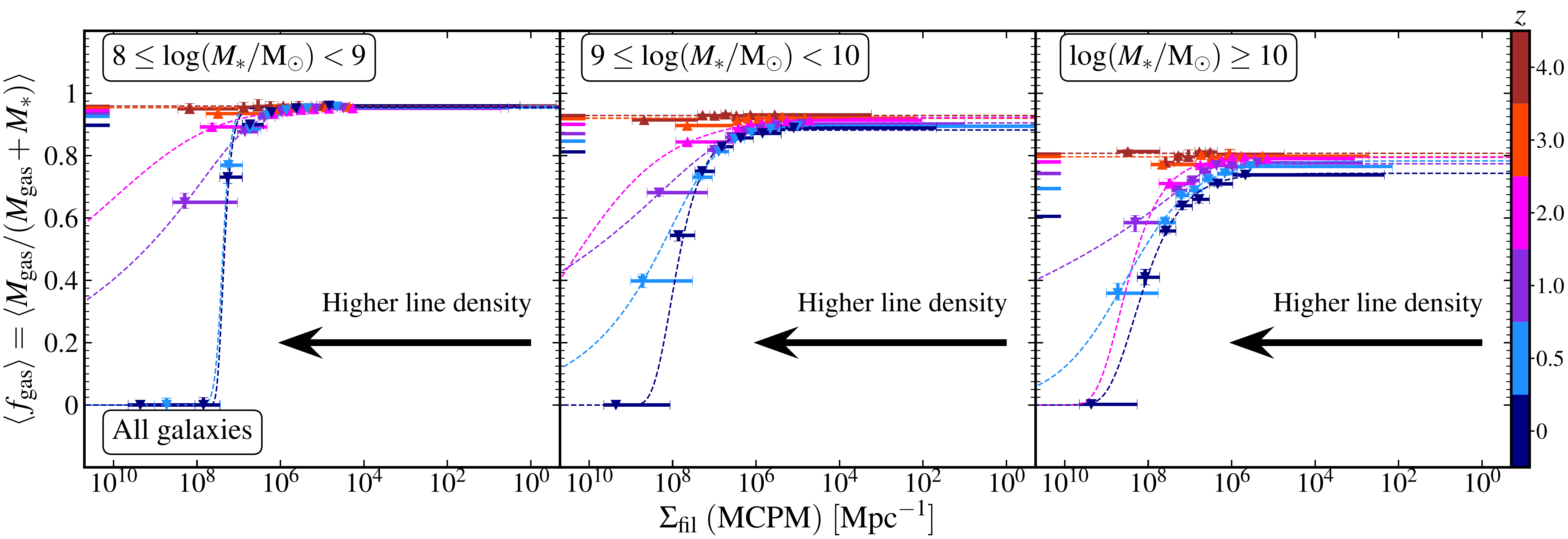}{0.88\textwidth}{}
} \vspace{-25pt}
\gridline{\fig{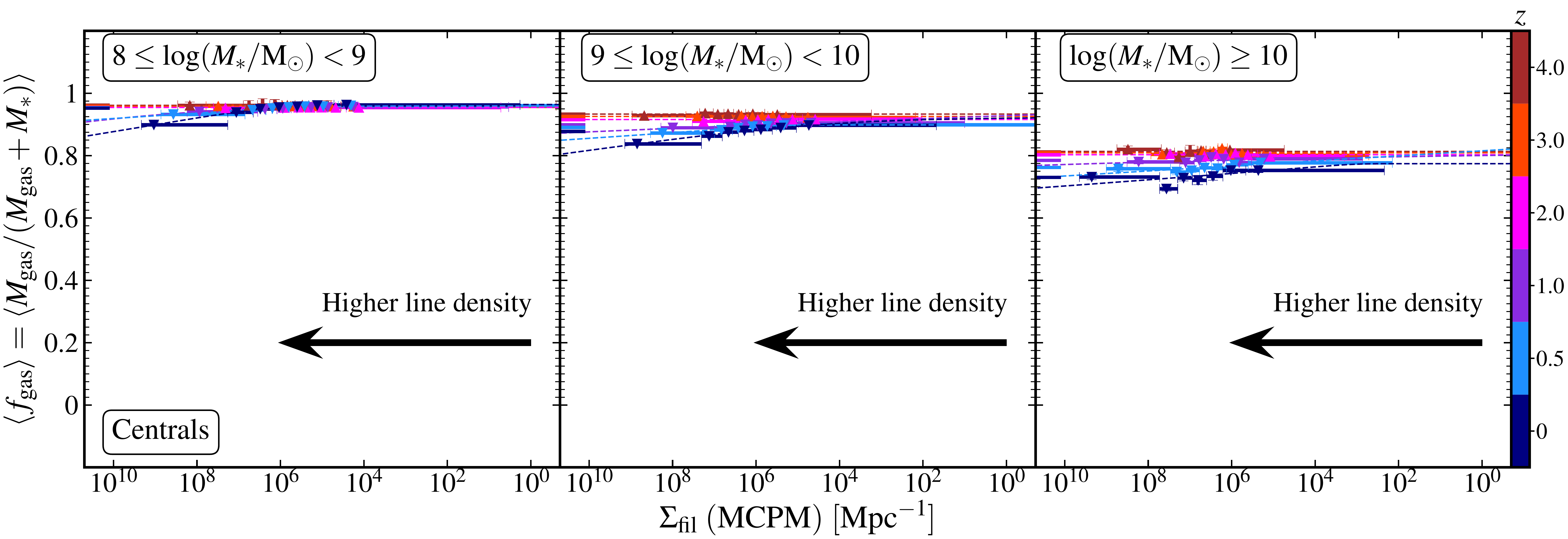}{0.88\textwidth}{}
} \vspace{-25pt}
\gridline{\fig{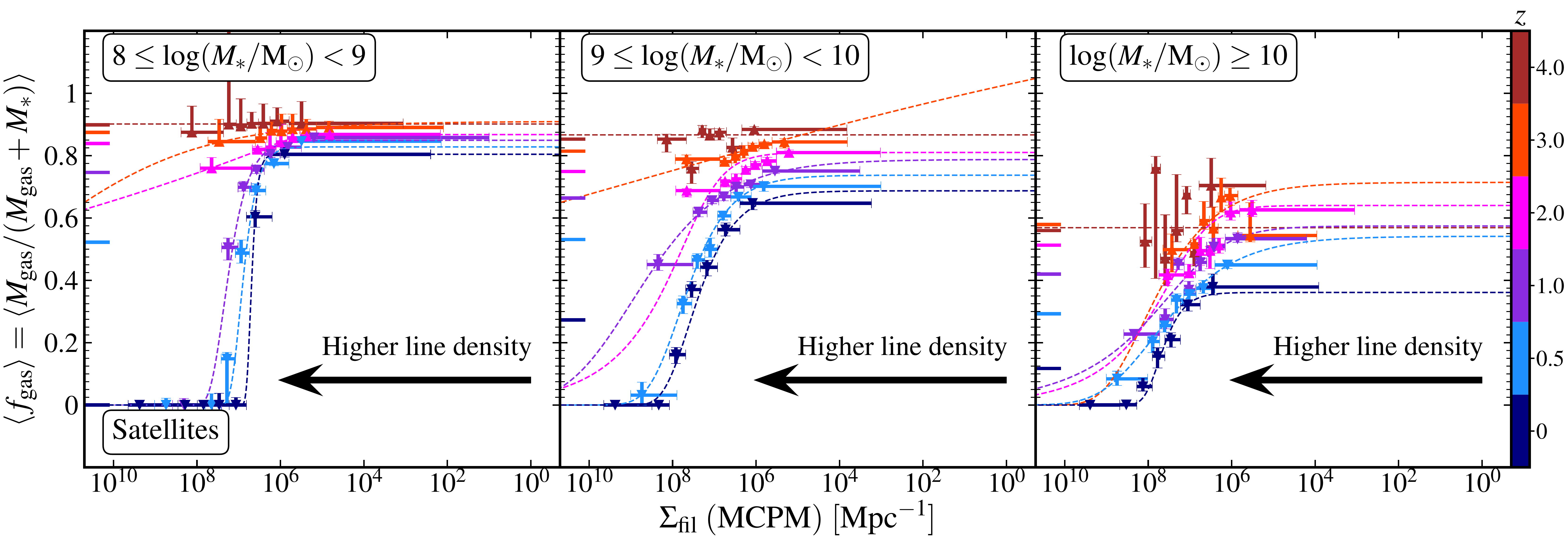}{0.88\textwidth}{}
} \vspace{-20pt}
\caption{
Similar to Figure~\ref{fig:medssfr_sigmafil}, but showing the {\medfgas}-{\sigmafil} relationship. The average gas fraction correlates strongly with the filament line density at lower redshifts, but the correlation is mostly driven by satellites. {\medfgas} in centrals is hardly {\sigmafil}-dependent, despite {\medssfr} in centrals showing such a dependence (Figure~\ref{fig:medssfr_sigmafil}).
}
\label{fig:medfgas_sigmafil}
\end{figure*}

The effect of filaments on quenching star formation in galaxies is complex, with many potential physical mechanisms at play. Here, we consider one of the more straightforward explanations: the availability of gas varies with the cosmic web environment. To this end, we measure the bound gas fraction of a galaxy, ${\fgas}={\mgas}/({\mgas}+{\ms})$, similar to \citetalias{Hasan23}, and study its dependencies on the distance and density of the nearest filaments of the cosmic web. {\fgas} contains all gas particles gravitationally bound to a halo and can be considered the gas fraction of the circumgalactic medium (CGM) and the interstellar medium (ISM). 
We recognize that in order to truly draw a direct causal link between gas supply and subsequent star formation in galaxies, we need to consider specifically the cold and dense gas that facilitates star formation. We defer a comprehensive analysis of the dependence of particular gas phases on the filamentary environment to a future study and instead simply consider below the cosmic web-dependence of all gas bound to a galaxy.

We measure {\medfgas} in terms of {\sigmafil} and present the results in Figure~\ref{fig:medfgas_sigmafil}. As in Figure~\ref{fig:medssfr_sigmafil}, we present the relationship {\medfgas}-{\sigmafil} for all galaxies (top), central galaxies (middle) and satellite galaxies (bottom) for three different mass ranges and six different redshifts. 
The data points and error bars represent the medians and bootstrapped errors, respectively, and the dashed curves represent the best-fit relationships of the functional form in Eq.~\ref{eq:betamodel}, where now $Y={\medfgas}$ and $Y_0$ is the value of {\medfgas} at $X_0$.

There is a pronounced decline in the median gas fraction with increasing filament line density for all masses at redshifts $z\leq1$. {\medfgas} drops by at least a factor of two from ${\sigmafil}\sim10^6~\mathrm{Mpc}^{-1}$ to $\sim10^9~\mathrm{Mpc}^{-1}$ at this epoch. At $z=0$, there is very little gas on average in galaxies near the highest filament line densities, which explains the corresponding lack of star formation in these regions. However, this is caused by satellite galaxies, whose gas fractions strongly depend on {\sigmafil} for all masses. On the other hand, the gas fractions of centrals are almost independent of {\sigmafil} across mass and redshift.

Interestingly, even at $z=2$, there is a small but noticeable reduction of gas at higher line densities. This is in contrast to the unchanging median star formation with increasing filament line density at $z=2$ (Figure~\ref{fig:medssfr_sigmafil}).
Furthermore, we saw a noticeable decline with {\sigmafil} in {\medssfr} of centrals of high mass at $z\leq0.5$ in Figure~\ref{fig:medssfr_sigmafil} (middle row), which is not mirrored in the dependence of {\medfgas} on {\sigmafil}.
We also find that in high-mass satellite galaxies, the average gas fraction tends to drop with high filament line density even out to $z\sim3$, which is not commensurate with average star formation staying constant with line density for these galaxies at these epochs. Nevertheless, we again caution against drawing general conclusions based on the low-number statistics of massive satellites at high $z$.
The pieces of evidence above {\it could} point to a differential star formation efficiency in galaxies depending on the filamentary environment. However, there are many other physical factors at play that could be responsible for these trends, including but not limited to, AGN feedback \citep[e.g.,][]{Fabian12}, cold flows at high $z$ \citep[e.g.,][]{DB06}, and recycling of gas in the CGM \citep[e.g.,][]{Muratov17}.
We defer an analysis constraining the relative importance of these effects in shaping the gas supply to future work.

\begin{figure*}
\vspace{-10pt}
\centering
\gridline{\fig{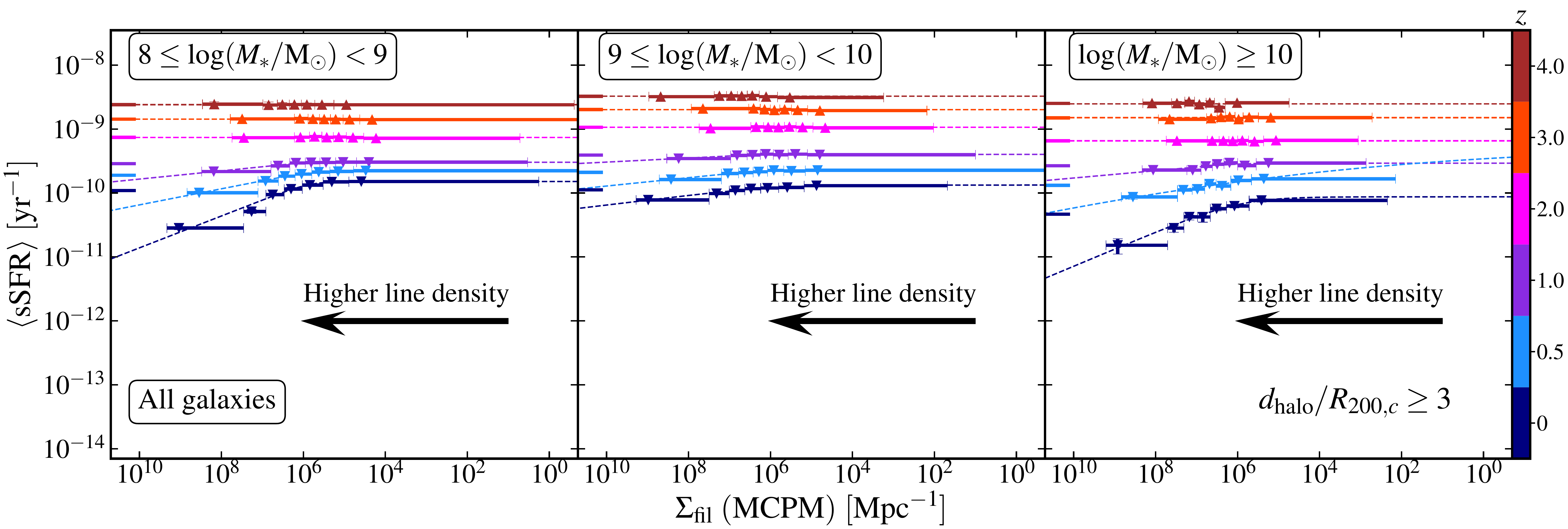}{0.86\textwidth}{}
} \vspace{-25pt}
\gridline{\fig{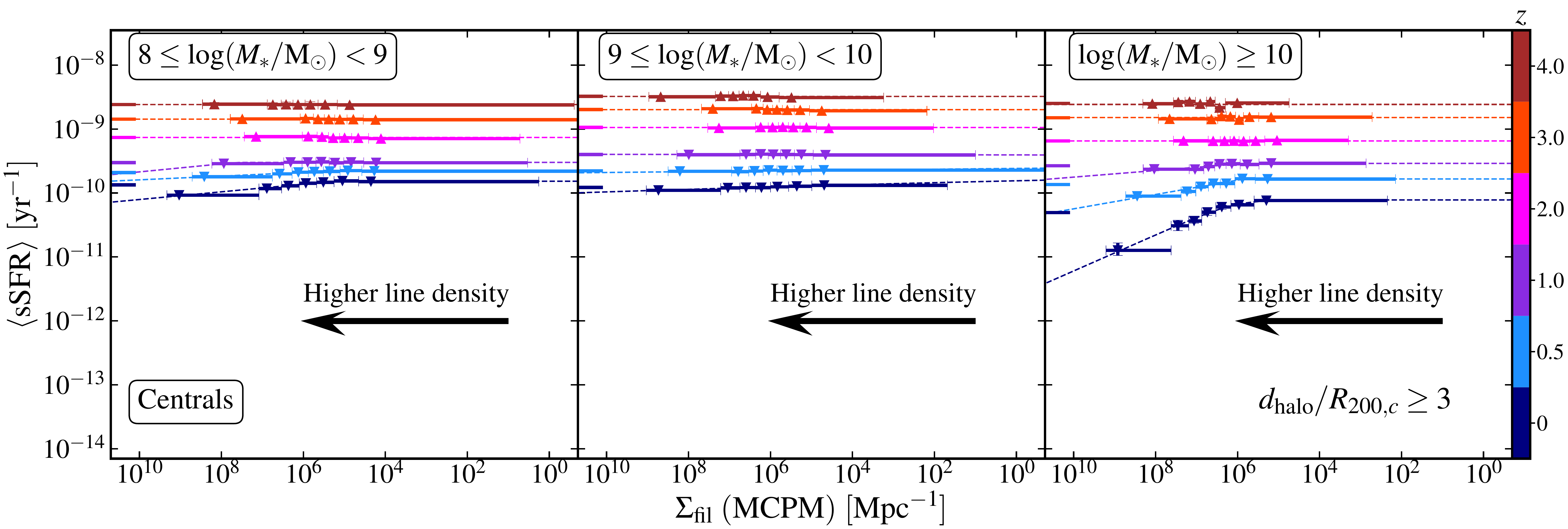}{0.86\textwidth}{}
} \vspace{-25pt}
\gridline{\fig{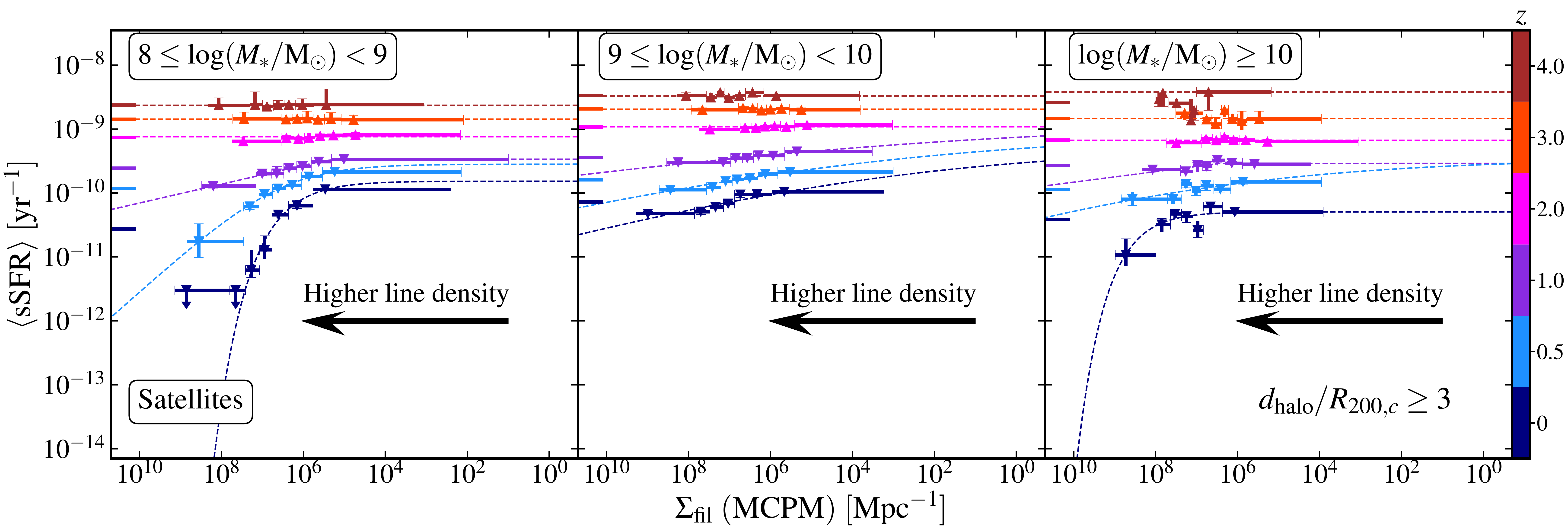}{0.86\textwidth}{}
} \vspace{-20pt}
\caption{
Same as Figure~\ref{fig:medssfr_sigmafil}, but now removing all galaxies within 3${\Rh}$ of the 0.5\% most massive halos at each redshift. Most of the dependence of quenching on high filament line density at low redshift persists even after removing the effect of massive halos.
}
\label{fig:medssfr_sigmafil_dhalo>3}
\end{figure*}

We can conclude from this section that the dependence of star formation activity on filament line density can be explained in most cases by the corresponding dependence on the total gas content of galaxies.
This is generally true for the entire galaxy population and is generally driven by satellites, with the exception of massive satellites at $z\geq2$. 
Central galaxies do not experience a significant change in gas fraction at higher filament line densities, despite a sizable drop in star formation, at later times.

\vspace{-5pt}

\subsection{Accounting For Massive Halos}
\label{sec:massive}

\begin{figure*}
\vspace{-10pt}
\centering
\gridline{\fig{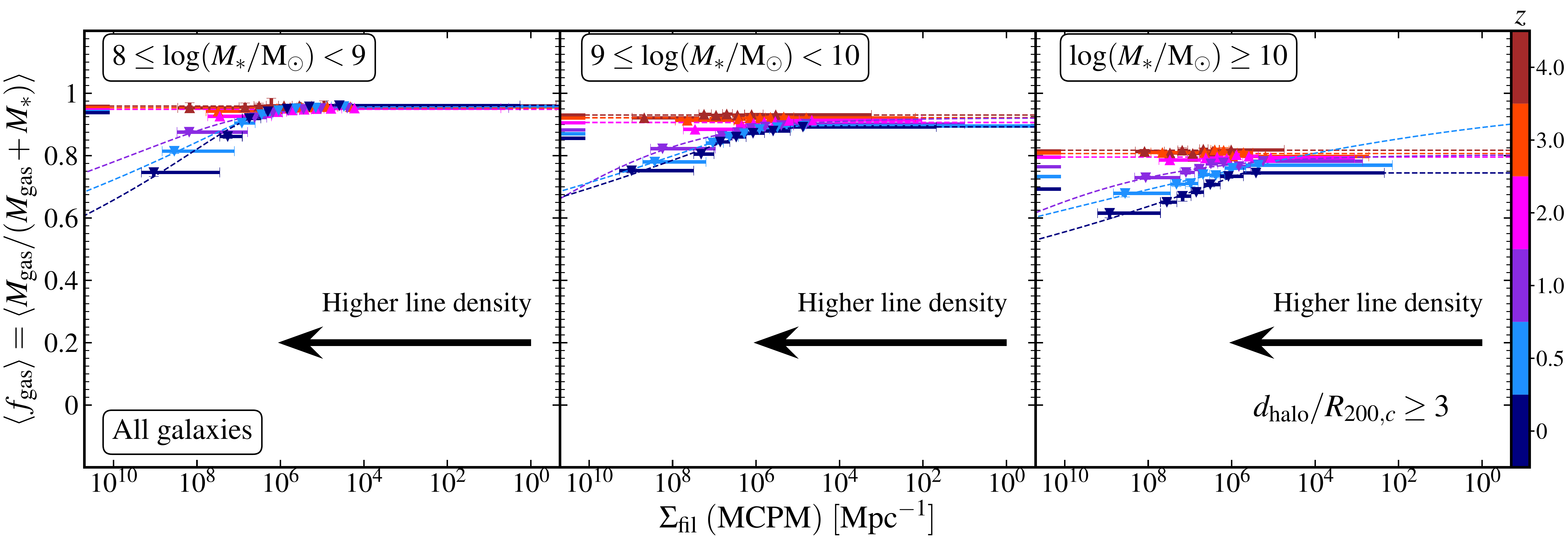}{0.85\textwidth}{}
} \vspace{-25pt}
\gridline{\fig{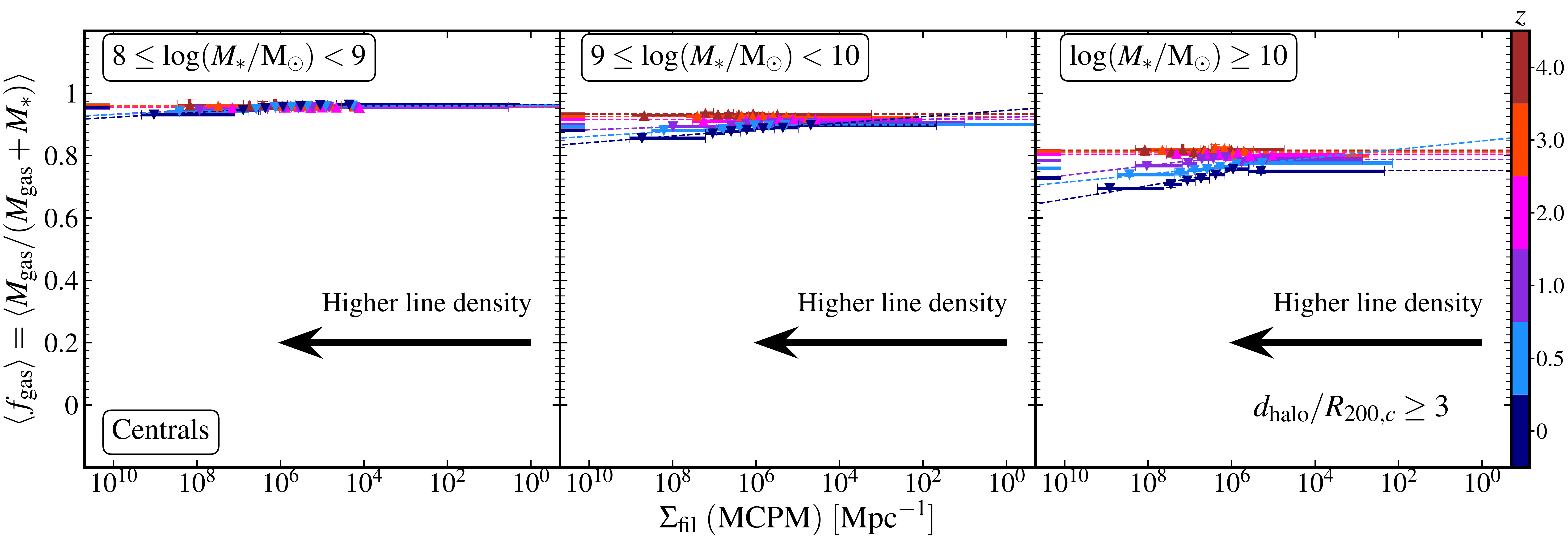}{0.85\textwidth}{}
} \vspace{-25pt}
\gridline{\fig{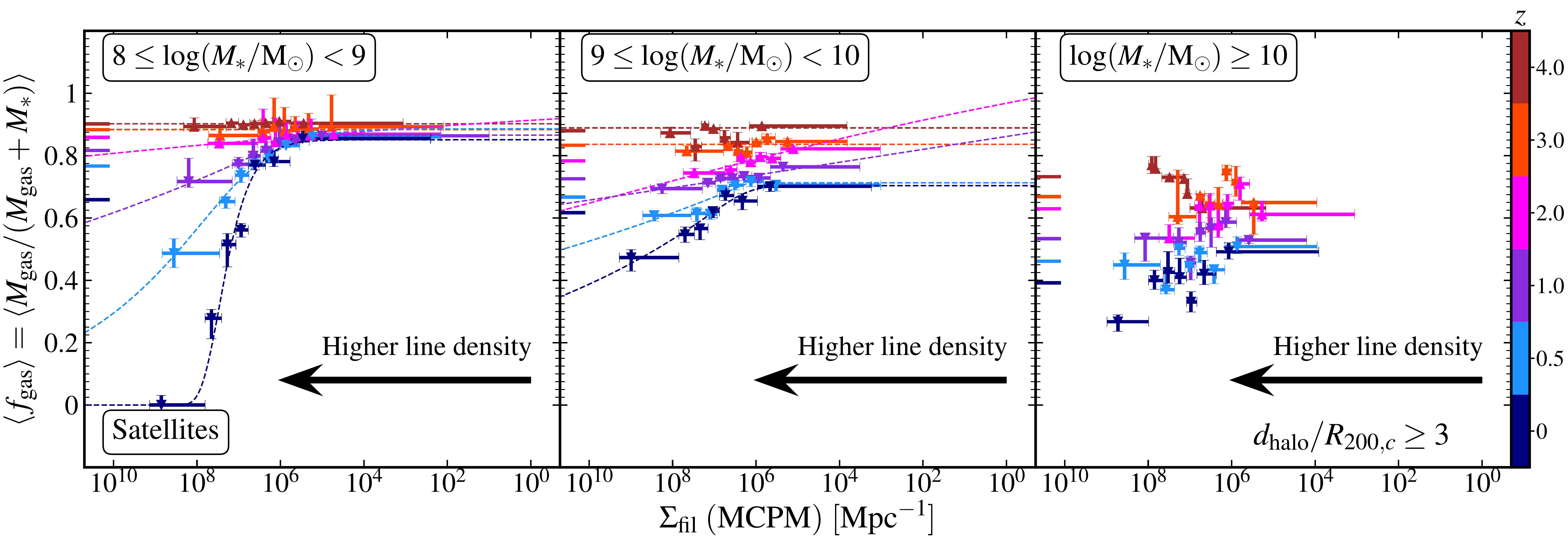}{0.85\textwidth}{}
} \vspace{-20pt}
\caption{
Same as Figure~\ref{fig:medfgas_sigmafil}, but now removing all galaxies within 3${\Rh}$ of the 0.5\% most massive halos at each redshift.
Gas supply is largely dependent on high filament line density at low redshift even after the effect of massive halos is removed.
}
\label{fig:medfgas_sigmafil_dhalo>3}
\end{figure*}

We appreciate the possibility that some or much of the dependence of galactic properties on large-scale cosmic web environments could actually stem from the corresponding dependence on local halo environment. We attempt to separate the effect of the small-scale environment (halo scales, up to several hundred kpc) from that of filaments ($>$Mpc) by controlling for the effect of massive halos in deriving the average relationship of star formation and gas fraction with {\sigmafil}. 
We estimate the median star formation and gas fraction relationships by now removing all galaxies within 3${\Rh}$ of the most massive halos at a given redshift. The halos we remove are the 0.5\% most massive halos at a given $z$ to be consistent with section~\ref{sec:identify}. This corresponds to ${\mh}\!\simeq\!1.4\!\times\!10^{13}~{\Msun}$ at $z=0$ to ${\mh}\!\simeq\!1.2\!\times\!10^{12}~{\Msun}$ at $z=4$.
Thus, we remove the contribution of nodes hosting group and cluster-mass halos at lower $z$ to find the residual effect of local filament line density beyond just the hydrodynamical effect of these environments.

In Figure~\ref{fig:medssfr_sigmafil_dhalo>3}, we present the {\medssfr}-{\sigmafil} relationships as in Figure~\ref{fig:medssfr_sigmafil}, having removed galaxies $<3{\Rh}$ of the 0.5\% most massive halos at a given $z$.
We find that much of the dependence of star formation on filament line density is preserved even after removing the effect of massive halos.
Galaxies of all masses experience at least a small drop in {\medssfr} relative to the median near high line density filaments at $z=0$. For low- and high-mass galaxies, there is a reduction in {\medssfr} of a factor of at least 2 from the lowest to the highest {\sigmafil} bin at $z\leq0.5$.
This effect is again driven mostly by satellites. The fact that low-mass satellites are quenched at $z\leq0.5$ at high line densities implies that the strong dependence of quenching on filamentary environment in this population is not merely due to the local environments of group/cluster-sized halos and their massive centrals.
High-mass satellites are also quenched in high-line density environments far outside the most massive halos.

The small dependence of {\medssfr} of low- and intermediate-mass centrals on {\sigmafil} at low $z$ (Figure~\ref{fig:medfgas_sigmafil}) mostly disappears when removing the effect of the most massive halos. We interpret this as the effect of massive halos neighboring lower mass centrals causing a slight reduction in sSFR. However, the star formation activity in high-mass centrals continues to show a dependence on {\sigmafil} even after removing these massive halos. For example, there is a $>$1 dex drop in {\medssfr} from  ${\sigmafil}\sim10^5~\mathrm{Mpc}^{-1}$ to ${\sigmafil}\sim10^9~\mathrm{Mpc}^{-1}$.
It seems that high-mass galaxies are least affected by the exclusion of galaxies close to massive halos.

Finally, in Figure~\ref{fig:medfgas_sigmafil_dhalo>3}, we present similar plots of the {\medfgas}-{\sigmafil} relationship at different {\ms} and $z$ after removing all galaxies within 3${\Rh}$ of the most massive halos. 
Note that for the high-mass satellites (lower right panel), we do not include best-fit relationships, as the functional form in Eq.~\ref{eq:betamodel} does not approximate the trends very well (most likely due to the small number of galaxies in each bin).
Similar to the above, our main finding is that massive halos cannot fully account for the dependence of average gas fraction on filament line density. At $z\leq1$, the drop in {\medfgas} at high {\sigmafil} does diminish significantly, but does not disappear entirely. For any galaxy mass at $z=0$, there is a non-negligible decline in {\medfgas} of $\approx$20\% from the lowest to the highest {\sigmafil} bin.
Once again centrals exhibit very little dependence on {\sigmafil}, while satellites show a strong dependence for all masses at lower $z$. The gas supply of low-mass satellites far from the most massive halos at $z\leq0.5$ dwindles considerably in high-line density filaments, which is consistent with their corresponding decline in star formation in these environments.
Interestingly, the slight dependence of {\medfgas} on {\sigmafil} at $z=$2--3 that we showed in Figure~\ref{fig:medfgas_sigmafil} disappears when the massive halo effect is removed. 
Note that the low-number statistics for high-mass satellites prevents us from drawing any particular conclusions about this population.

We also perform these experiments with a more conservative halo mass cut of the top 1\% of the halo masses rather than the top 0.5\%. We find that, while there is some small reduction in the degree to which filament line density correlates with suppressing gas supply and star formation, the change is not drastic. Our qualitative takeaway -- that the drop in {\medssfr} and {\medfgas} at high {\sigmafil} at later times persists after removing galaxies within 3${\Rh}$ of massive halos -- does not change.

The results in this section imply that the decrease in gas supply and star formation of many galaxies with increasing filament line density at low redshift cannot be explained as simply a local effect stemming from the most massive halos.
Also, once we remove galaxies close to the most massive halos, there is no dependence of either average star formation or average gas fraction on filament line density at $z\geq2$. 
In closing, we remark that the local environment will inevitably be linked to the global cosmic web environment, as more massive halos and regions of higher local density will be in nodes and high-line density filaments. Hence, disentangling the effect of local density from that of filaments is not trivial.


\section{Discussion}
\label{sec:discuss}


\subsection{Improved Cosmic Web Reconstruction}
\label{sec:improve}

We showed that 
cosmic filaments are identified with much higher fidelity when using the MCPM density field in {\disperse} instead of the prepackaged DTFE density field. 
First, much lower persistence structures are identified, revealing an entire population of low-line density filaments that the DTFE-based method misses (Figure~\ref{fig:vis_fils}). 
The same regions of space that would be considered voids from the DTFE method, but that are somewhat overdense regions of the mass distribution, are shown to contain a small number of low-line density, diffuse filaments with the MCPM method.  
Second, the morphologies of MCPM 
filaments are more physically plausible  than that of the DTFE filaments, with rounded curves instead of sharp line segments.
Third, from MCPM, we retrieve a much smaller fraction of short, likely spurious, filaments close to the smoothing scale of the input density field (${\Lfil}\lesssim1$~Mpc) than from DTFE (Figures~\ref{fig:cal} and \ref{fig:lfil}). Overall, MCPM filaments are probing much larger physical structures than DTFE filaments.
Finally, comparing MCPM and DTFE density fields smoothed on the same resolution scale results in MCPM mapping significantly better to the true underlying matter distribution (from DM) with far less scatter than DTFE (Figure~\ref{fig:densitycomp}).

There are several potential reasons why the MCPM density field drastically improves upon filament identification. The primary reason is the difference in sparsity of the DTFE and MCPM density fields, where the latter is a continuous estimate at all locations of the input volume (i.e., every $(0.5~\mathrm{Mpc})^3$ for $256^3$ resolution) and the former only provides density estimates at the locations of galaxies (i.e., $\approx1/10000$ of the information contained in the MCPM density field) and linearly interpolates the density elsewhere. Effectively, this means that the {\it mse} structure extractor in {\disperse} has a much more complete and accurate density field from which to compute gradients, local minima, maxima, and saddle points, which allows the topological identification scheme to recover underlying structures more faithfully. 
We generated continuous mass-normalized DTFE density fields independent of {\disperse} and found that these do not trace the DM overdensity as well as MCPM fields, especially in low-density regions. 
Unlike DTFE, MCPM traces the underlying matter density field across a wide dynamic range 
within even diminutive structures.

The DTFE method, while widely used and applicable to many different types of input data, is known to be highly susceptible to noise in small point set data such as galaxy locations as opposed to large point sets such as simulation particle data
\citep[e.g.,][]{RD07,Pandey13,BW20}. In a comparative study of different density field estimators applied to SDSS and simulation data, \citet{Ferdosi11} found that DTFE both underestimates the true density in underdense regions and substantially overestimates the true density in high-density regions.
We comment that 
even the continuous DTFE density field estimation is local linear.
DTFE is robust and unbiased, but that means it can be significantly underutilizing its input information (if there is supplementary physical information it is ignoring). 
MCPM, on the other hand, is a non-linear density estimator, and it can be physically calibrated.

Previous applications of the DTFE/{\disperse} method have found that using individual particles as tracers in simulations (as opposed to, e.g., galaxies) can increase the number of faint structures detected. \citep[e.g.,][]{Kuchner20,Kotecha22}.
Galaxies are biased tracers of the cosmic mass distribution. \citet{Zakharova23} showed that, using DM particles as inputs in {\disperse} {\it mse} results in more of the filaments being found than using galaxies as inputs.
They also showed that filaments identified from DM tracers are, on average, longer than those identified from galaxy tracers. 
We reran the {\disperse} filament identification using DTFE density estimation from DM particles as tracers instead of galaxies. The resulting outputs, depending on the adopted persistence threshold, show filamentary structures similar to those of our MCPM density field run on galaxies as tracers. We verified through both visual and statistical comparisons that the output structures generated from applying DM particles as tracers in DTFE and galaxy positions as tracers in MCPM produce similar qualitative results. In other words, the conclusions of our paper do not change between density estimation with DTFE using DM particles and that with MCPM using galaxies.

However, in the real universe, we do not have access to a continuous DM mass distribution, which traces the {\it true} cosmic matter distribution. Instead, we can only access sparser tracers, such as spectroscopically surveyed galaxies.
Our method employs MCPM density fields and does not require a large number of particles as tracers. A relatively sparse catalog of galaxies is enough to identify the highly detailed filamentary structure that traces the underlying matter distribution with remarkable accuracy. This significantly reduces the storage and computational requirements for identifying less prominent filaments of the cosmic web (a $z=0$ snapshot of TNG100 is 1.7 TB in size), 
and has substantial implications for identifying filaments from any observed galaxy catalog (see Section~\ref{sec:obs}).

\vspace{-5pt}

\subsection{Filaments Affecting Quenching and Gas Supply?}
\label{sec:fileffect}

The new filament finding method using MCPM and {\disperse} that we introduced in this work reveals a wide range of filaments that the traditional DTFE-based method misses. This includes many low-persistence filaments in less dense regions of the universe than high-persistence filaments but are, nevertheless, diffuse bridges of matter connecting galaxies. Some authors have posited the existence of ``tendrils'' or thin filaments, which are small structures embedded within voids \citep{Alpaslan14b,CroneOdekon:2018aa,Porter23}.
Most of these studies classify tendrils as intermediate-density environments between filaments and voids, finding varying degrees of difference in the star formation activity and gas supply of galaxies in these structures. \citet{CroneOdekon:2018aa} showed that galaxies in the ALFALFA survey \citep{giovanelli05} in tendrils are on average more massive but similar in color and {\HI} gas fraction to those in voids, while filament galaxies are less gas-rich and redder. Similar results were reported for the GAMA survey \citep{Driver11} by \citet{Alpaslan16}.

The cosmic web filaments we identify from the MCPM density fields reveal a population of tendril-like low-persistence filaments. When considering the distance of galaxies to all filaments identified by our new method, there is generally very little dependence of galactic star formation or gas fraction on this distance -- a consequence of the fact that most galaxies live within $\sim$1.5 Mpc from a filament spine. However, there is a very notable dependence of galactic properties on the local 1D filament line density we define (Eq.~\ref{eq:sigmafil}). 
Interpreting galaxies living near low-line density filaments as tendrils, we see that these are the galaxies with highest star formation activity and gas fraction at any given mass and redshift (Figs.~\ref{fig:medssfr_sigmafil} and \ref{fig:medfgas_sigmafil}), consistent with the results of \citet{CroneOdekon:2018aa} and \citet{Alpaslan14b}. Therefore, a key result of our work is the discovery of a {\it differential effect} of filaments based on their line density; galaxies have less gas and star formation in higher line density filaments, but galaxies are the most gas-rich and star-forming in lower line density filaments.

\vspace{-5pt}

\subsubsection{Galaxy-Filament Phenomenology Across Cosmic Time}

\citet{Zhu22} reconstructed the cosmic web using a Hessian-based method in a cosmological simulation run with the adaptive mesh refinement (AMR) code {\sc RAMSES} \citep{teyssier02} and studied the gas accretion rates in DM halos residing in tenuous/thin (with diameters $<$3 Mpc~$h^{-1}$) and prominent/thick (diameters $>$3 Mpc~$h^{-1}$) filaments. They found that the gas accretion rate to ${{\logmh}<12}$ halos is noticeably lower when they reside in thicker filaments -- by $\approx$20-30\% at $z=0.5$ and $\approx$200-300\% at $z=0$. At higher $z$, they found no significant differences in gas accretion rates between thin and thick filaments in higher-mass halos. They also reported a minimal dependence of the gas accretion rate on the distance from the filament spine to the halo center. 
Although we quantified filament line density instead of width, our results are broadly consistent with theirs, since the availability of gas in the halo ({\fgas}) and star formation activity in the galaxy (sSFR) should generally follow from the rate at which gas accretes onto the halo. Namely, galaxies in dense filaments have much lower gas fraction and star formation activity than those in diffuse filaments, an effect that increases with time (decreasing redshift). 
\citet{Zhu22} demonstrated that there is significantly more hot ($T>10^6$~K) gas in thicker filaments at lower redshift, which can result in the gas supply to galaxies, e g., via cold accretion streams \citep{Keres09}, in these regions being cut off.  
Recently, \citet{Lu23} identified quantified filamentary boundaries by the gas density profiles, baryon fraction, the existence of a shock, and virial equilibrium, all of which are consistent.

Numerous theoretical explorations have been made on the distribution of gas in cosmic web filaments at low redshift \citep[e.g.,][]{GE21,GE22,Gouin22} and high redshift \citep[e.g.,][]{Mandelker21,Ramsoy21,Lu23}, whose results have far-reaching implications for the impact of filaments on galaxy formation. 
Using a cosmological zoom-in simulation run with the {\sc AREPO} code as used in TNG, \citet{Lu23} demonstrated that there are distinct radial zones, characterized by their own gas thermodynamic properties, in filaments at $z\sim4$. 
They showed that filament cores contain cool, low-entropy gas shielded from the shock-heated exteriors, interpreting this zone as thin cold streams that are long believed to be the primary sources of feeding high-$z$ galaxies \citep[e.g.,][]{DB06,Pichon11}. 
The high-resolution simulation of \citet{LG19} run with the {\sc GADGET-3} code \citep{Springel05} showed that filaments can accelerate gas cooling and subsequent star formation inside halos at $z=4$ and $z=2.5$.

We note that the physical scales in the studies mentioned above vary from kpc-scale filaments to Mpc-scale filaments, which are the focus of this study. \citet{AC07} devised a method to identify cosmic web structure at multiple spatial scales by smoothing a DTFE density field on the scales of interest.
Leveraging a different method, \citet{Tempel14} identified filaments in SDSS on scales of order $0.5h^{-1}$ Mpc. The MCPM density fields we generate in this work are smoothed on scales close to $\approx$0.5 Mpc, thus rendering any identified filaments on sub-Mpc scales (a small fraction of our catalogs) poorly resolved and likely unphysical. The findings we report herein, therefore, pertain only to the effect of large-scale filaments across cosmic time. Smaller-scale filaments are not captured by the resolution of the density field. In contrast, the high-$z$ filaments characterized by studies such as \citet{Mandelker20}, \citet{Ramsoy21} and \citet{Lu23} are more like narrow coherent streams, of order 1--100 kpc in length, that branch out of large-scale filaments as they penetrate DM halos.

In this work, we found that the gas fraction and the star formation activity were at similar levels near and far from MCPM filaments and near high- and low-line density filaments at the early epochs of $z\geq2$. This provides evidence that filaments can efficiently supply gas into galaxies at early times (perhaps owing to thin cold streams), which allows them to grow quickly.  Our results suggest that even filaments of comparable line density (measured in terms of comoving filament segment length) are more effective in channeling gas into galaxies and fueling star formation at $z\gtrsim2$ than at $z<2$. 
Some authors, such as \citet{Birnboim16}, showed that filamentary condensation could occur for unstable filaments at early times, allowing the thin streams mentioned above to feed halos with cold gas.
If indeed cold cores are ubiquitous in filaments at high-$z$, regardless of the filament line density, then this would result in similar gas supply levels and subsequent star formation in galaxies located near almost all types of filaments. 
This might, at face value, be interpreted as a lack of filamentary dependence of gas supply and star formation at high-$z$. 
A lack of strong dependence of star formation activity and/or galactic gas supply on environment 
before the so-called ``cosmic noon'' of star formation activity \citep[e.g.,][]{MD14} is a result that is being corroborated by a plethora of recent observational \citep[e.g.,][]{Darvish16,Moutard18,Momose22,Shi24} and theoretical  \citep[e.g.,][]{Xu20,Chang22,Malavasi22,Bulichi23} work.

On the other hand, we do see a strong dependence of galactic phenomena on the cosmic web at later times. 
As filaments accrete matter from underdense voids, they grow in density, and the physical properties of gas evolve toward less hospitable conditions for cold gas supply and star formation. 
In a cosmological zoom-in simulation using {\sc Ramses}, \citet{Ramsoy21} found that the filament gas density profile can be modeled as an isothermal cylinder with a core radius that closely tracks the size of the galaxy in which the filament ends, and evolving as $\sim\!(1+z)^{-2.7}$. 
However, once again, we note the different physical scales of filaments in their study and ours.
We wish to explore the gaseous properties of filaments using simple analytical models in future work.

\vspace{-5pt}

\subsubsection{Physical Mechanisms at Play}

In line with expectations from a long line of theoretical studies, \citet{Martizzi19} found that in TNG100, there is a strong increase in the amount of hotter and denser gas in filaments from high $z$ to the local universe, gradually resulting in the formation of a pervasive warm-hot intergalactic medium \citep[WHIM; e.g.,][]{Dave01,CO06}. 
With time, there is an increase, therefore, in physical mechanisms that can cut off the gas supply and/or quench galaxies living in denser filaments. One such mechanism is the formation of accretion shocks at the boundaries of filaments and other cosmic web structures, which can suppress gas accretion and ultimately quench galaxies at high-$z$ \citep[e.g.,][]{Pasha22} and low-$z$ \citep[e.g.,][]{Zinger18,Li23}. ``Cosmic web stripping'', a form of ram pressure stripping of gas near filamentary cores, has been invoked by \citet{BL13} to explain how local dwarf satellites can lose their gas. This mechanism is in line with our finding in this work and in \citetalias{Hasan23} that satellites have strongly suppressed star formation and gas fraction at low $z$ in dense cosmic environments. This is in contrast to the slow removal of gas, often referred to as ``starvation'' or ``strangulation'' which can cut off gas supply and quench galaxies on relatively long timescales on the order of a Gyr \citep[e.g.,][]{Larson80,Peng15}. 

Many of these processes are encapsulated by the ``cosmic web detachment'' model of \citet{AC19}, which explains the increase in quenching and decrease in gas content close to the cosmic web as a natural consequence of galaxies being detached from their primordial cold gas-supplying filaments (or streams) at later times. 
In particular, our finding that satellites are strongly quenched in high-line density filaments could be an effect of these galaxies being increasingly cut off from primordial filaments with time. 
We also note that it is also possible for galaxies to be ``pre-processed'' -- i.e., experience environment-driven quenching -- before accreting onto filaments \citep[e.g.,][]{Fujita04,LG19}.
While a detailed analysis of gas physical conditions such as temperature, density, and entropy is beyond the scope of this work, this is a topic we intend to explore to understand the physical connection between filamentary gas and galaxy formation at different epochs. 
In future work, we also aim to predict directly measurable gas properties such as column density and kinematics, which can be compared to QSO absorption-line observations \citep[e.g.,][]{Burchett16,Hasan20}, as a function of cosmic web environment.

In this work, we also quantified the relative effects of galaxy mass and filamentary environment on quenching galaxies in the local universe. We repeat an analysis similar to that of \citet{Peng10} but using the filament line density as an environmental metric. We showed that {\sigmafil} is an effective quantity for assessing the environmental impact of the fraction of quenched galaxies. 
At $z\!\!=\!\!0$, ${\logms}\!\!\gtrsim\!\!10.5$ galaxies are almost all quenched, regardless of the filament they live near or whether they are centrals or satellites. 
This is mostly a consequence of the black hole (BH) feedback model of TNG, where the growth of a central BH to mass $\log(M_{\mathrm{BH}}/{\Msun})\!>\!8.2$ results in a very rapid transition of the host galaxy to a quiescent state \citep{Terrazas20}. The onset of a low-accretion ``kinetic'' mode of BH feedback consisting of a BH-driven wind in the TNG model is known to transform the thermodynamics of gas in the CGM and quench star formation in galaxies with ${\logms}\!\gtrsim\!10.5$ \citep{Zinger20}. This result does create some tension with the observations \citep[e.g.,][]{Donnari21b}.
However, AGN feedback is widely believed to be among the most important factors in quenching massive galaxies at any redshift \citep[e.g.,][]{DiMatteo05,Fabian12,SD15}.

This is in line with the picture of {\it mass quenching} advocated by \citet{Peng10} and others to show the dominant role that internal mechanisms such as stellar and AGN feedback play in regulating star formation in massive galaxies. At lower masses, however, the filamentary environment plays an important role in quenching galaxies. High line density filaments have more quenched and red galaxies with ${\logms}<10.5$ than low line density filaments. Like \citet{Peng12}, we find that this {\it environmental quenching} is dominated by satellite galaxies, albeit our work considers the global cosmic web environment rather than the local galaxy overdensity.  
We also showed that the most massive halos are not solely responsible for the galaxy-filament connection (Figures~\ref{fig:medssfr_sigmafil_dhalo>3} and \ref{fig:medfgas_sigmafil_dhalo>3}) at low-$z$, thus necessitating an environmental effect of large-scale structure beyond the hydrodynamics of massive halos such as groups and cluster.

One of our findings is the lack of correspondence between the filamentary dependence of sSFR and gas fraction in some cases. 
While satellite galaxies generally lose their gas supplies and become passive systems as they approach high-line density filaments in later times,
high-mass low-$z$ centrals experience a somewhat significant reduction in {\medssfr}, but not in {\medfgas}. 
Some physical mechanisms that may result in the reduction of star formation efficiency -- loosely referring to the ratio of star formation to gas fraction -- in high filament line densities environments are (a) hydrodynamical conditions of gas in high line density filaments, e.g., higher temperatures and pressures, are inhospitable to star formation, (b) internal mechanisms such as AGN feedback, heats up gas in the CGM and prevent star formation as a result \citep[e.g.,][]{Voit15,Oppenheimer20,DV22}, and (c) gas in the recycled within the halo after accretion \citep[e.g.,][]{AA17,Muratov17}. 
Crucially, however, we do not constrain the fraction of cold and dense gas which is a more direct measure of star formation fuel.
We defer a detailed analysis of the cosmic web dependence of cold/dense gas, BH activity, and/or halo gas recycling to future work.

Finally, we make some remarks about numerical and modeling considerations in interpreting our results on the galaxy-cosmic web connection. One such consideration is the impact of feedback in the gas fraction and star formation of galaxies. \citet{Dave20} compared the gas in various cosmological simulations with different subgrid prescriptions and numerical methods. They found that despite broad agreement between different simulation suites in global statistical distributions and scaling relations (such as the gas-to-stellar mass relation),  there are quantitative differences in galaxy properties. The differing model of AGN and supernovae (SNe) feedback between TNG and EAGLE, for example, causes much lower {\HI} gas fraction in low-mass galaxies in EAGLE. However, TNG is also known to somewhat overproduce gas fractions in massive galaxies relative to observations \citep[e.g.,][]{Diemer19,Stevens19}. Numerical resolution is also a source of uncertainty which will no doubt affect the gas supply, and ultimately quenching, of galaxies \citep[e.g.,][]{Pillepich19,Nelson20}. 
Lastly, the baryonic mass resolution of $\sim\!1.4 \!\times\! 10^{6}~{\Msun}$ in TNG100 means that the lowest mass galaxies we study, i.e. ${\logms}=8$, might not be very well-resolved, and therefore the statistics of low-mass galaxies ({\lowmsrange}) carry some numerical uncertainty.

\vspace{-5pt}

\subsection{Application to Observations}
\label{sec:obs}

We have demonstrated that this new cosmic web filament-finding methodology can be applied to any catalog of galaxies to extract highly detailed information on the characteristics of filaments and their effect on galaxy evolution. 
We are conducting a {\it Hubble Space Telescope} (HST) archival research program
(AR\#17568; PI: Hasan) to apply this method to observed galaxy catalogs from SDSS
at $z\simeq0.5$ \citep{Wilde23}.
We will probe the gas in filaments across $\approx$40\% of the age of the universe using absorption features in QSO spectra observed with the Cosmic Origins Spectrograph (COS) on board HST \citep{Green12}.

We expect to apply our method to other state-of-the-art observational surveys of galaxies at even higher redshifts. Large surveys with sky coverage similar to SDSS from the Dark Energy Spectroscopic Instrument (DESI) 
\citep[see][and references therein]{Lan:2023_desiVisual}, will allow the identification of cosmic web filaments out to $z\simeq1.6$. 
The Subaru Prime Focus Spectrograph (PFS)
Galaxy Evolution program will observe up to half a million galaxies at redshifts $0.7 \!\lesssim\! z \!\lesssim\! 7$ , including a deep survey of $>10^5$ continuum-selected galaxies out to $z\sim2$ \citep{Greene22}. In addition, the Euclid Wide Survey \citep{Euclid22} and the SPHEREx All-sky Survey \citep{Dore16} will provide the large galaxy catalogs needed to reconstruct the cosmic web over most of the cosmic time. Such datasets will enable us to identify filaments from MCPM cosmic density fields at the peak epoch of star formation activity and test our predictions of a lack of cosmic web dependence of quenching and gas supply of galaxies in the first $\approx$4 Gyr of the universe.

Additionally, the {\it James Webb Space Telescope} (JWST), thanks to programs such as JADES \citep{Eisenstein23}, will provide galaxy spectra at the epoch of cosmic dawn ($z\gtrsim4$) over much smaller fields of view. This will allow us to generate limited cosmic web reconstructions in the first $\sim$1 Gyr of the universe. 
Ultimately, our filament finding method can be applied to galaxy catalogs observed by the Roman Space Telescope, especially the planned High Latitude Wide Area Survey, which will observe millions of galaxies up to $z\sim3$ \citep{Wang:2022_romanHLSS}.



\section{Conclusion}
\label{sec:conclusion}

In our study, we used the {\disperse} algorithm to identify filaments of the cosmic web in the TNG100 simulation at redshifts $z\!=\!0$, 0.5, 1, 2, 3, and 4. We analyzed the output filamentary structure and its impact on galaxy evolution using two different density field estimates, the prepackaged DTFE density field \citep{SV00} and the MCPM density field inspired by {\it slime mold} \citep{Burchett20}. Our findings are summarized below.

\begin{enumerate}
\itemsep-2pt
\item {\disperse} more accurately identifies filaments when using the MCPM density field than the DTFE density field. Filaments identified by the MCPM method are more numerous, provide a more accurate representation of the cosmic mass distribution, and have more natural shapes, than those identified by the DTFE method. Additionally, the MCPM method can identify lower-density, less prominent filaments that the DTFE method misses.
These results are a consequence of MCPM tracing the DM density field with higher accuracy and less scatter than DTFE.

\item $\gtrsim$90\% of galaxies are located within a distance of $\approx$1.5 Mpc from the spine of an MCPM filament. On the other hand, many galaxies reside far away from DTFE filaments. As a result, the median sSFR {\medssfr} 
shows minimal dependency on the distance to the filament spine when MCPM filaments are taken into account, unlike in the case of DTFE filaments.

\item We define the 1D local line density of filaments, denoted {\sigmafil}, as the total MCPM overdensity per unit length along a filament segment. At low redshifts ($z\leq1$), star formation is suppressed in galaxies near high-line density filaments. However, the correlation is almost non-existent at higher redshifts ($z\geq2$). High-line density filaments strongly quench satellites of all masses, while high-mass ({\highmsrange}) centrals also have lower star formation in these environments at lower redshifts.

\item At a redshift of $z=0$, we find that the fraction of galaxies that are quenched (${\log(\mathrm{sSFR}/\mathrm{yr}^{-1})\!<\!-11}$) and red (${g-r\!>\!0.6}$) are not affected by their environment when their stellar mass is above ${{\logms}>10.5}$. However, at lower masses, galaxies tend to be redder and more likely to be quenched when the filament line density is higher for a given galaxy mass. Additionally, satellites are far more susceptible to the effects of the filamentary environment than central galaxies.

\item Most of the dependence of {\medssfr} and {\medfgas} on filament line density at low $z$ persists even removing all galaxies within 3${\Rh}$ of the most massive halos, implying that these effects cannot be explained by massive halos halo. We conclude that the line density of filaments plays a significant role in forming stars via the gas supply available to galaxies. 
The filament line density and matter overdensity at which galaxies begin to quench depends on galaxy mass; lower mass galaxies quench at lower {\sigmafil} than higher mass galaxies up to a stellar mass of ${\logms}\!\sim\!10.5$.

\end{enumerate}

\vspace{-5pt}

Our scientific results point to a picture in which high-density filaments of the cosmic web help reduce the gas supply and star formation activity of galaxies in the last $\sim$10~Gyr of cosmic time. 
In the early universe, thin, narrow streams efficiently feed galaxies even within denser filaments, allowing ubiquitously high levels of star formation activity.
Galaxies are likely increasingly detached from their cold gas supply due to physical mechanisms such as gas stripping or shock-heating from filaments, leading to quenching.
While satellites are most susceptible to the effect of high-line density filaments, centrals also experience a reduction in star formation at later times. 
Hydrodynamical effects of gas in filaments likely play a role in regulating galactic gas supply and star formation, but internal effects such as AGN feedback are far more dominant in high-mass galaxies.

We hope that our work will be the first of many to take advantage of this density estimation method to identify cosmic web filaments. Similar reconstructions can be performed on other hydrodynamical cosmological simulations such as EAGLE \citep{Schaye15}, SIMBA \citep{Dave19}, and FIREbox \citep{Feldmann23}, as well as different resolution and volume runs of the TNG project, including TNG50 \citep{Pillepich19} and MilleniumTNG \citep{Pakmor23},
to extract important insights into the effect of subgrid physics, stellar/AGN feedback, numerical resolution, etc. on the properties of large-scale structures and their interplay with galaxies.
Our new approach can be applied to rich observational datasets of galaxy catalogs from state-of-the-art current and upcoming observatories such as SDSS, DESI, Subaru PFS, JWST, Euclid, SPHEREx, and Roman.
This has the potential to unlock a rich array of investigations in many research areas of extragalactic astrophysics and cosmology.


\vspace{-10pt}

\section*{Data Availability}

The catalogs used in this work are available on Zenodo under an MIT license: 
\dataset[10.5281/zenodo.10962244]{https://doi.org/10.5281/zenodo.10962244}. 

\vspace{-12pt}

\begin{acknowledgements}

The authors are grateful to N. Luber and Z. Edwards for their help setting up {\disperse}.
We acknowledge insightful discussions at the 2023 Santa Cruz Galaxy Workshop. 
We also acknowledge stimulating discussions with V. Pandya, H. Aung, H. Zhang, J. Powell, C. Cadiou, L. Scholz-D{\'i}az, S. Simha, K-G. Lee, R. Momose, F. van den Bosch, N. Gluck, C. Pichon, U. Kuchner, C. Welker, M. Strauss, R. Dav{\'e}, J. Greene, R. Teyssier, N. Luber, J. van Gorkom, X. Wang, B. Frye, S. Juneau, S. Alberts, S. Tonnesen, L. Sales, D. French, J. Tumlinson, J. Wu, P. Mansfield, K. Finlator, A. Coil, and S. Alberts.
This research used resources of the National Energy Research Scientific Computing Center, a DOE Office of Science User Facility supported by the Office of Science of the US Department of Energy under Contract No. DE-AC02-05CH11231 using the NERSC award HEP-ERCAP0024028.
FH and JNB are supported by the National Science Foundation (NSF) LEAPS-MPS award $\#2137452$. 
OE is supported by an incubator fellowship from the Open Source Program Office at UC Santa Cruz funded by the Alfred P. Sloan Foundation (G-2021-16957).
DN is supported by the NSF AST-2307280 grant.
NM acknowledges support from the Israel Science Foundation (ISF) grant 3061/21 and US-Israel Binational Science Foundation (BSF) grant 2020302.

\end{acknowledgements}



\bibliographystyle{aasjournal_nikki}
{\tiny \bibliography{Refs}}

\end{document}